\documentclass[onecollarge]{svjour2}       % onecolumn "king-size"
\smartqed  % flush right qed marks, e.g. at end of proof
\usepackage{graphicx}
\usepackage{epsfig,multirow,amsmath,enumerate}
\usepackage{hyperref}
\usepackage{graphicx}
\usepackage{float}
\usepackage{color}
\usepackage{amsfonts}
\usepackage{amssymb}
\usepackage{natbib}
\usepackage[a4paper,margin=2.5cm]{geometry}

\journalname{arxiv preprint}
\newcommand\beq{\begin{equation}}      % begin equation numbered
\newcommand\beqnn{\begin{eqnarray*}}   % begin equation no number
\newcommand\beqa{\begin{eqnarray}}     % begin equation array
\newcommand\beqann{\begin{eqnarray*}}  % begin equation array no number
\newcommand\eeq{\end{equation}}        % end equation numbered
\newcommand\eeqnn{\end{eqnarray*}}     % end equation no number
\newcommand\eeqa{\end{eqnarray}}       % end equation array numbered
\newcommand\eeqann{\end{eqnarray*}}    % end equation array no number
\newcommand{\notimplies}{%
  \mathrel{{\ooalign{\hidewidth$\not\phantom{=}$\hidewidth\cr$\implies$}}}}

\newcommand{\mb}[1]{\mathbf{#1}}                                                            % mathbf
\newcommand{\ave}[1]{\langle #1 \rangle}
\newcommand{\ket}[1]{\left| #1 \right\rangle}                                               % ket
\newcommand{\bra}[1]{\left\langle #1\right|}                                               % bra
\newcommand{\overlap}[2]{\left\langle {#1} | {#2} \right\rangle}                            % <A|B>
\newcommand{\matrel}[3]{\left\langle {#1} \right| {#2} \left| {#3}\right\rangle}            % <A|B|C>
\newcommand\bi{\begin{itemize}}
\newcommand\ei{\end{itemize}}
\newcommand{\nl}{\nonumber \\}
\newcommand{\eref}[1]{(\ref{#1})}
\newcommand{\sref}[1]{Section~\ref{#1}}

\newcommand{\Eref}[1]{Equation (\ref{#1})}

\newcommand\etal{\emph{et al. }}
\begin{document}

\title{Operator formalism for topology-conserving crossing dynamics in planar knot diagrams%\thanks{Grants or other notes
}
\subtitle{}


\author{C.M. Rohwer \and
        K.K. M\"uller-Nedebock %etc.
}


\institute{C.M. Rohwer \at
              Institute of Theoretical Physics, University of Stellenbosch, Stellenbosch 7600, South Africa \\
              \email{crohwer@sun.ac.za, crohwer@is.mpg.de}             \\
	      \emph{Present address:} Max-Planck-Institut f\"ur Intelligente Systeme, Heisenbergstraße 3,
D-70569 Stuttgart, Germany %  if needed
           \and
           K.K. M\"uller-Nedebock \at
Institute of Theoretical Physics, University of Stellenbosch, Stellenbosch 7600, South Africa
}

\date{\today}

\maketitle

\begin{abstract}
We address here the topological equivalence of knots through the so-called Reidemeister moves. These topology-conserving manipulations are recast into dynamical rules on the crossings of knot diagrams. This is presented in terms of a simple graphical representation related to the Gauss code of knots. Drawing on techniques for reaction-diffusion systems, we then develop didactically an operator formalism wherein these rules for crossing dynamics are encoded. {The aim is to develop new tools for studying dynamical behaviour and regimes in the presence of topology conservation}. This necessitates the introduction of composite paulionic operators. The formalism is applied to calculate some differential equations for {the time evolution} of densities and correlators of crossings, subject to topology-conserving stochastic dynamics. {We consider here the simplified situation of two-dimensional knot projections. However, we hope that this is a first valuable step towards} addressing a number of important questions regarding the role of topological constraints {and specifically of topology conservation} in dynamics through a variety of solution and approximation schemes. Further applicability arises in the context of the simulated annealing of knots. The methods presented here depart significantly from the invariant-based path integral descriptions often applied in polymer systems, {and, in our view, offer a fresh perspective on} the conservation of topological states and topological equivalence in knots. 

\end{abstract}

%%%%%%%%%%%%%%%%%%%%%%%%%%%%%%%%%%%%%%%%%%%%%%%%%%%%%%%%%%%%%%%%%%%%%%%%%%%%%%%%%%%%%%%%%%%%%%%%%%%%%%%%%%%%
%%%%%%%%%%%%%%%%%%%%%%%%%%%%%%%%%%%%%%%%%%%%%%%%%%%%%%%%%%%%%%%%%%%%%%%%%%%%%%%%%%%%%%%%%%%%%%%%%%%%%%%%%%%%
%%%%%%%%%%%%%%%%%%%%%%%%%%%%%%%%%%%%%%%%%%%%%%%%%%%%%%%%%%%%%%%%%%%%%%%%%%%%%%%%%%%%%%%%%%%%%%%%%%%%%%%%%%%%
%%%%%%%%%%%%%%%%%%%%%%%%%%%%%%%%%%%%%%%%%%%%%%%%%%%%%%%%%%%%%%%%%%%%%%%%%%%%%%%%%%%%%%%%%%%%%%%%%%%%%%%%%%%%
\section{Introduction}
\label{intro}

In this article we address topological constraints related to entanglements in a statistical physical setting. Entanglements occur naturally, for instance in the setting of polymer systems, where they impose constraints on conformational freedom. Several factors contribute to (and impose) topological constraints. Firstly, \emph{strands physically cannot move through each other}. This self-exclusion property is not trivial to incorporate into mathematical polymer descriptions. (In path integral formulations of excluded volume interactions in flexible polymers, for instance, a perturbation expansion in terms of the excluded volume parameter diverges when treated in fewer than four spatial dimensions. The divergence occurs since monomer contacts for random walks are unlikely for dimensions above four, but grow as a function of polymer length for dimensions below four \cite{vilgis2000polymer,rubinstein2003polymer}. Other descriptions of self-avoiding walks (SAWs) exist. Simple examples include Flory's basic scaling arguments for the entropy and energy of SAWs \cite{rubinstein2003polymer}, and mapping the excluded volume interaction onto the $n \to 0$ limit of $n$-component spin model \cite{degennes1979scaling}.)

Secondly, we need to consider how topological constraints are affected by whether \emph{strands are open or closed}. Closed 
%polymer
loops have ``frozen in'' topological constraints that are absent for open strands. For open strands, the relative entanglement of the strands is not a fixed property of a particular configuration since strands may slide along each other -- a process known as reptation \cite{degennes1979scaling} in poylmer physics -- until entanglements disappear. In contrast, the conformational freedom in closed loops depends greatly on whether loops are interlocking or not. Individual closed loops can be self-entangled, and therefore also have frozen-in topological constraints that are absent for open strands. (In the setting of polymers, for instance, further types of frozen-in topological states may exist, e.g. in cross-linked polymer networks \cite{deamedwards1976,duplantier1986polymer}.)

Here we shall consider individual closed loops, viewed mathematically as knots. There exist several biological systems where closed polymer loops appear naturally. A suited example is that of ring closure observed in DNA molecules, which has been the subject of much theoretical and experimental study; see, for instance \cite{shore1981dna}, where it was concluded that ``short DNA fragments are surprisingly flexible'' and that ``covalent joining of the ends of linear DNAs by ligase to form closed circular molecules is a fast reaction''. Several experimental observations in biological systems, many of them pertaining to coiling states DNA strands, have been addressed through topological approaches \cite{dewitt}. In a cellular context the role of topology is highlighted by the requirement for compact packaging of DNA strands. Indeed, specialised topoisomerase enzymes regulate winding properties of DNA by altering its topological states \cite{thompson2008}. On the other hand, developments in experimental techniques have made possible the synthesis of molecular knots and a direct study of their topological properties \cite{molecknots}.

{
The theoretical understanding of the role of topological constraints has, not surprisingly, been the focus of considerable research efforts. Monographs by Kauffman \cite{kauffman} and Nechaev \cite{nechaev1999} and a review article by Kholodenko and Vilgis \cite{kholodenko1998}, amongst others, provide a wide range of background from different perspectives.  The last two references are particularly relevant for questions of entanglements of polymers. 
}

{
The preservation of the knotted state of polymers is a reduction of their configurational freedom in comparison to ghost-like chains (i.e. which move through one another). The physical picture of such a constrained polymer is amenable to very useful \emph{effective} descriptions such as models restricting the chains to tubes (for example, \cite{rubinstein2003polymer,doiedwards}); but there are also very successful path-integral implementations of the actual topological restriction of other objects on a closed polymer loop (see, e.g. \cite{Rostiashvili1993}).  The effective tube descriptions do not, however, capture the topology of polymer knots completely, although they are mathematically much simpler to implement.  Historically, topological constraints on closed polymers have typically been addressed in the context of knot theory, with the goal of classifying knots which possess some common topological property or knot invariant. Given two knots, $K_1$ and $K_2$, a sensibly defined knot invariant $I(K_i)$ should allow for some conclusions about the topological equivalence of $K_1$ and $K_2$. Several knot invariants have been defined on knot diagrams (planar projections of knots) --- see, for instance, \cite{kauffman,gilbert,lickorish}.  Examples include simple numbers like winding or linking numbers, and polynomial invariants like the Jones and Alexander polynomials.  However, as yet it is uncertain whether any invariant provides a complete classification scheme for knots. Alexander polynomials, for instance, do not distinguish between all types of knots. Jones and Kauffman polynomials provide a more powerful classification scheme, but do not distinguish all knot types either \cite{huang1996}. Algebraic invariants and other algebraic approaches have been applied extensively to the problem of knot simplification and statistical properties of knots, see e.g. works by Grosberg and Nechaev \cite{nechaev1999,nechaev1996,grosberg1992algebraic}. In application to polymer systems with closed loops, then, typically some knot invariant $I(K)$ is included into polymer path integrals through a delta functional which is then exponentiated through the Fourier representation. In this way some aspects of topology conservation are captured by restricting the conformations for the partition function.  These mathematical descriptions are complicated as even using the simplest knot invariants to determine the partition function of a constrained polymer system is a non-trivial matter.  In a seminal article by Edwards \cite{edwards1967}, a closed polymer wound around a rod is investigated through the use of winding numbers as a knot invariant. This problem has been revisited and extended \cite{grosbergfrish2003,rohwer2014} to include geometric confinement.  The references \cite{nechaev1999,kholodenko1998} discuss and point to a wealth of physics connections, applications and implementations of knots, including field theories.
}

In the present work, {which is inspired by the above themes from polymer physics}, we take a slightly different approach: instead of asking whether knots $K_1$ and $K_2$ have a common knot invariant, \textbf{\emph{we focus on the conservation of the topological state of a particular knot}}.
%investigate whether it is possible for $K_1$ to be \textit{manipulated} into the form of $K_2$ (or vice-versa). 
Our aim is therefore not to find a comparative schema for knot classification through invariants, but rather to pursue a formalism that will allow us to {explore}
%generate all
knots that are topologically equivalent to a given knot. Our approach is based on the Reidemeister moves, which are fundamental to knot theory \cite{reidemeister} since they provide a necessary and sufficient recipe for manipulating knot diagrams in a way that leaves the knots' topology unchanged. Two knots that are related by any sequence of these moves are generally referred to as being regularly isotopic \cite{kauffman}. This is sufficient where we consider topology \textit{conservation} and no further classification scheme is needed.

The article is organised as follows. In Section \ref{knotchapt} we set out basic definitions of knots and describe how they may be represented in terms of planar projections. We address there the Reidemeister moves and their implications for topological equivalence. {Thereafter, we introduce an intuitive and instructive representation of knot diagrams (which we call the ``arc diagram'' representation) in terms of their crossing structure, not unlike the Gauss code. The topology-conserving Reidemeister moves are then cast into dynamical rules for these arc diagrams in Section \ref{bowdiagsect}. After reviewing some operator techniques for describing reaction-diffusion systems in Section \ref{doisectionmain}, we set out in Section \ref{doisectbow} an operator formalism for stochastic crossing dynamics on knot projections, subject to topology conservation. The aim there is to consider dynamics that evolve knot diagrams according to \textit{topology-conserving rules}. Similar techniques have been applied to studying various dynamical regimes and steady states in the presence of birth and decay processes, e.g. contact processes and pair-contact processes \cite{houch,paessensschuetz,baumann}. Our formalism extends these ideas to composite species with restricted occupation numbers, and allows for the systematic derivation of differential equations for the evolution of various densities and correlators for crossings of knot diagrams whose topological state is conserved, as is discussed in detail for the zeroth and first Reidemeister moves. This opens the door to a different perspective for investigating the role of topological constraints in dynamical processes. Applications and limitations of this operator formalism, as well as the connection to physical three-dimensional knots, will need to be studied in terms of specific examples, as set out in our outlook to future work.}  

{Although the question of knot simplification is not our focus, we present in Appendix \ref{simchapter} some suggestions toward an algorithm for simulated annealing of knots (knot simplification), incorporating the representation and dynamical rules from previous sections. }

%%%%%%%%%%%%%%%%%%%%%%%%%%%%%%%%%%%%%%%%%%%%%%%%%%%%%%%%%%%%%%%%%%%%%%%%%%%%%%%%%%%%%%%%%%%%%%%%%%%%%%%%%%%%
%%%%%%%%%%%%%%%%%%%%%%%%%%%%%%%%%%%%%%%%%%%%%%%%%%%%%%%%%%%%%%%%%%%%%%%%%%%%%%%%%%%%%%%%%%%%%%%%%%%%%%%%%%%%
%%%%%%%%%%%%%%%%%%%%%%%%%%%%%%%%%%%%%%%%%%%%%%%%%%%%%%%%%%%%%%%%%%%%%%%%%%%%%%%%%%%%%%%%%%%%%%%%%%%%%%%%%%%%
%%%%%%%%%%%%%%%%%%%%%%%%%%%%%%%%%%%%%%%%%%%%%%%%%%%%%%%%%%%%%%%%%%%%%%%%%%%%%%%%%%%%%%%%%%%%%%%%%%%%%%%%%%%%

\section{Knots, crossings and their representations}
\label{knotchapt}
As mentioned above, several motivations from biological, chemical and polymer systems exist to incorporate knot theory into theoretical polymer descriptions. Indeed, a closed, self-entangled polymer loop could be viewed as a knot. {We will be concerned simply with knot projections.} In this section we discuss some aspects of knots and their representations. We further address the Reidemeister moves --- local manipulations on planar knot diagrams that conserve the knot topology. In particular, we show how these moves may be translated into rules for dynamics of the crossings of knot diagrams.

\subsection{On knots and knot diagrams}
\label{whatisaknot}
We consider here only classical knots --- embeddings of a circle in three-dimensional Euclidean space. Virtual knots, which are a generalisation of standard knots \cite{kauffman2006virtual}, will not be dealt with. Physically such embeddings could be realised by taking an open piece of string, entangling it with itself in some chosen way, and closing it on itself. The closing captures (freezes) some aspects of the particular entanglement. Clearly different knots may be such that one cannot be continuously deformed into the other. This forms the basis for our notion of topological equivalence.

\begin{figure}[h]
	\centering
		\includegraphics[width=0.4\textwidth]{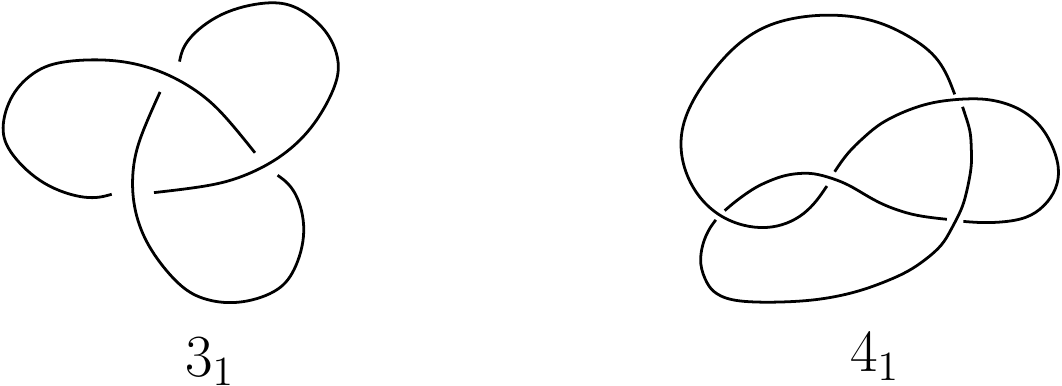}
	\caption{Knot diagrams for the knots $3_1$ (trefoil knot) and $4_1$. }
	\label{fig:knot4_1}
\end{figure}

The shadow of a knot is defined as the two-dimensional (i.e., planar) projection of the knot. We deal here with regular projections, for which the shadow is a regular graph with vertices that all have a degree of four. A knot diagram is then a shadow where some line-segments are deleted at the crossings to indicate over- or undercrossings; examples are shown in Figure \ref{fig:knot4_1}. Link diagrams are knot diagrams comprised of multiple components that do not intersect, but may be knotted / linked. We shall consider only individual knots, i.e., links with a single component.

Prime knots cannot be reduced or decomposed to simpler knots through manipulations that do not break strands, and are classified according to the number of crossings they contain. The knots $3_1$ and $4_1$ in Figure \ref{fig:knot4_1} are prime knots; they are in their ``simplest form'' in that the number of their crossings cannot be reduced through topology-conserving manipulations. Clearly these knots are topologically distinct. The question arises as to how topologically distinct knots may be classified. As stated, many knot invariants have been defined to this end, examples including winding numbers, linking numbers and several polynomial invariants; see, for instance \cite{gilbert, lickorish}. Alternatively one may ask in what way topologically equivalent knots are related. Indeed, it is this question that is of interest to us here. With that goal we now turn to a set of topological manipulation rules --- the Reidemeister moves.

\subsection{Reidemeister moves and knot equivalence}
\begin{figure}[h]
	\centering
		\includegraphics[width=0.6\textwidth]{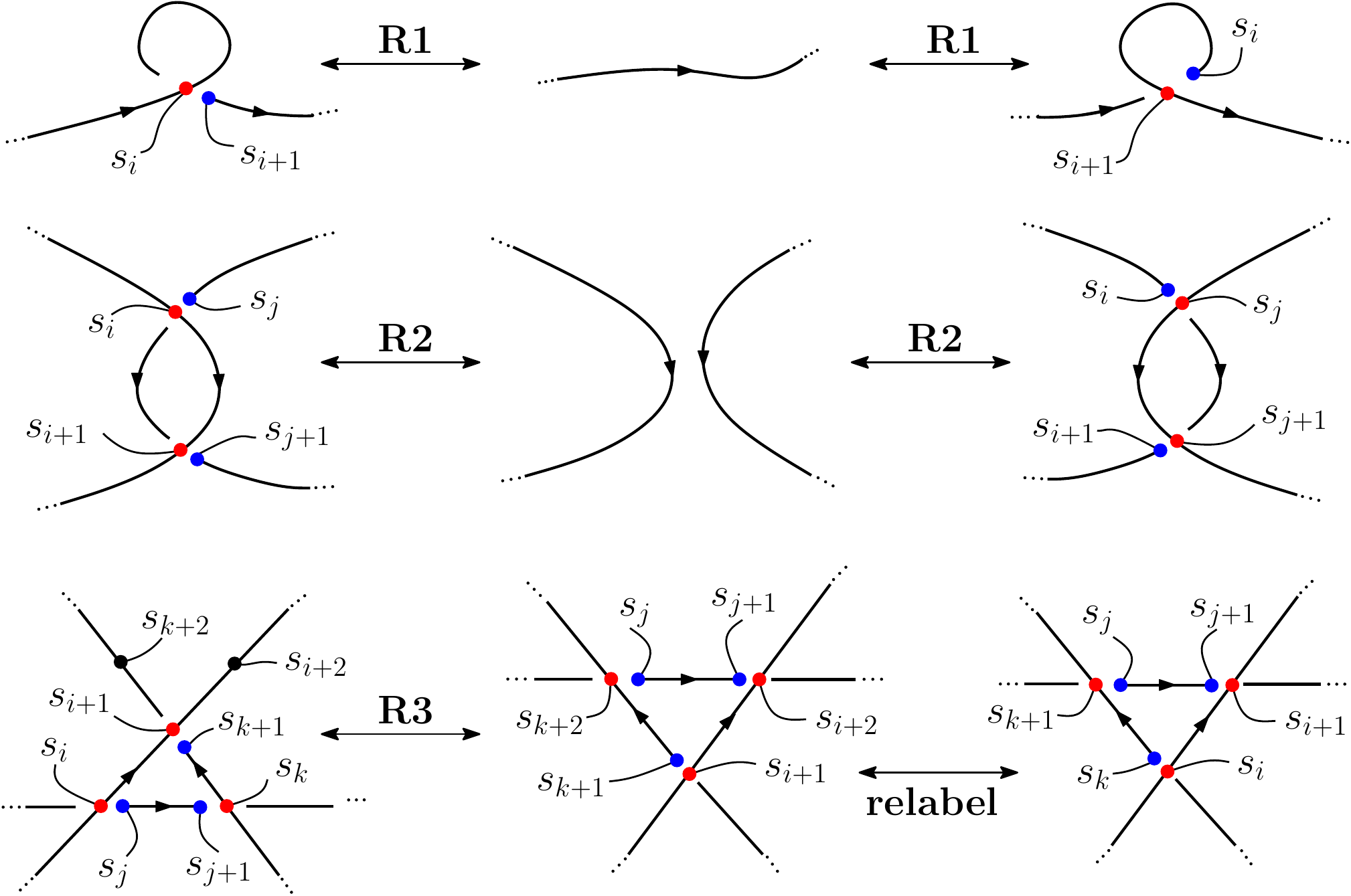}
	\caption{The Reidemeister moves on parts of some (unspecified) knot.}
	\label{fig:moveswithlabels}
\end{figure}

In his seminal work on knot theory \cite{reidemeister}, Reidemeister set out that rules for how knots or links may be manipulated without altering their topology. These manipulations, the Reidemeister moves, here denoted as \textbf{R1}, \textbf{R2} and \textbf{R3}, involve local manipulation of strands on a knot diagram, and are illustrated in Figure \ref{fig:moveswithlabels}. \textbf{R1} involves the removal of a single loop from a strand that is crossing with itself (or the addition of such a loop to a naked strand). \textbf{R2} entails the separation of two strands that cross each other in two places (or moving two separate strands so that they cross each other). Finally, \textbf{R3} involves moving one strand across a single crossing of two other strands. Naturally all three moves are reversible, and none forces strands to intersect. 

We have included in Figure \ref{fig:moveswithlabels} an orientation on the strands involved, and indicated discretised arc-length coordinates $s_i$ in the projection (relative to some starting point on the knot in question). Overpasses and underpasses are indicated with red and blue dots at the crossings, respectively. (The two dots at a given crossing are understood to be on top of each other in the planar projection.)

\begin{figure}[h]
	\centering
		\includegraphics[width=0.6\textwidth]{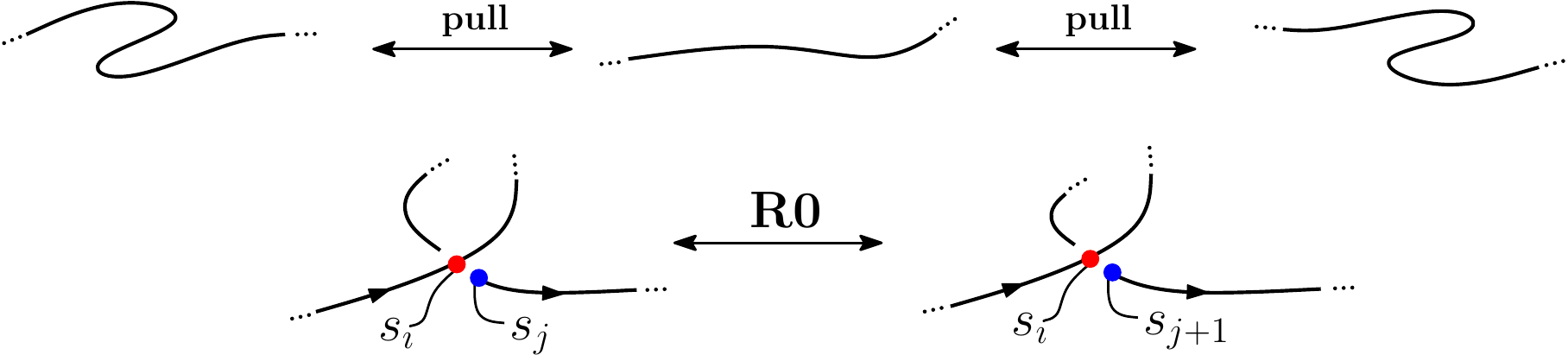}
	\caption{Trivial manipulations on knot diagrams.}
	\label{fig:movestrivial}
\end{figure}

A further move involves basic topological deformations of planar curves that do not alter the crossing structure of the knot. This move is topologically trivial, and may be viewed as stretching and pulling a knot without affecting its crossings \cite{gilbert}, as shown in the top of Figure \ref{fig:movestrivial}. Frequently this move is referred to as \textbf{R0} \cite{kauffman}. However, we shall reserve the notation \textbf{R0} for the topologically trivial move that alters the relative lengths of different segments between crossings by sliding strands across each other at a crossing in such a manner that no crossings disappear and no new crossings are introduced --- see the bottom scenario in Figure \ref{fig:movestrivial}. 

In his famous theorem, Reidemeister established that two knots are equivalent (isotopic) if and only if there exists some sequence of the Reidemeister moves that relates them \cite{reidemeister}. This notion will be used as the definition of topological equivalence of knots here.

In the following section we present a scheme for representing knot diagrams according to their crossings. This scheme will be used to derive rules on crossings of knots that encode the topology conservation captured by the Reidemeister moves.

\subsection{Crossings: allocation of signs and representation on arc diagrams}
We wish to record positions of crossings on a knot diagram. Restricting the discussion to tame knots with a finite length $L$, we introduce an arc length parameter $s \in [0,L]$, which describes the position relative to an arbitrary ``starting point'' in the knot diagram where $s = 0$ (and/or where $s = L$, since the knot is unbroken and periodic). We discretise $s$, so that $s_i = i \epsilon$ where $\epsilon=\frac{L}{N}$ would be a minimal length-scale / Kuhn length of the strands.

Signs are allocated to crossings on the knot diagram according to the following procedure,
\begin{enumerate}
\item Choose a reference point on the knot to be labelled as $s = 0$, and choose an orientation for the knot diagram (arbitrary).
\item Follow the strand until a crossing is encountered, say at $s_i$. Note the coordinate of the other strand involved with the crossing, say $s_j$.
\item If the current strand, $s_i$, is an overpass (red dot), record $\sigma({s_i,s_j})=+$. If the current strand is an underpass (blue dot), record $\sigma({s_i,s_j})=-$. The ordering $s_i<s_j$ is assumed.
\item On a line with sites representing the discretized coordinates, draw an arc connecting $s_i$ and $s_j$, labeled with $\sigma({s_i,s_j})$.
\item Continue in this manner until all crossings have been visited.
\end{enumerate}
Recall the usual assumption \cite{gilbert} that the projection is such that at most two strands lie above each other at a given crossing. This implies that a given site on an arc diagram be occupied by \textit{at most one arc foot}.

\begin{figure}[h]
	\centering
		\includegraphics[width=0.5\textwidth]{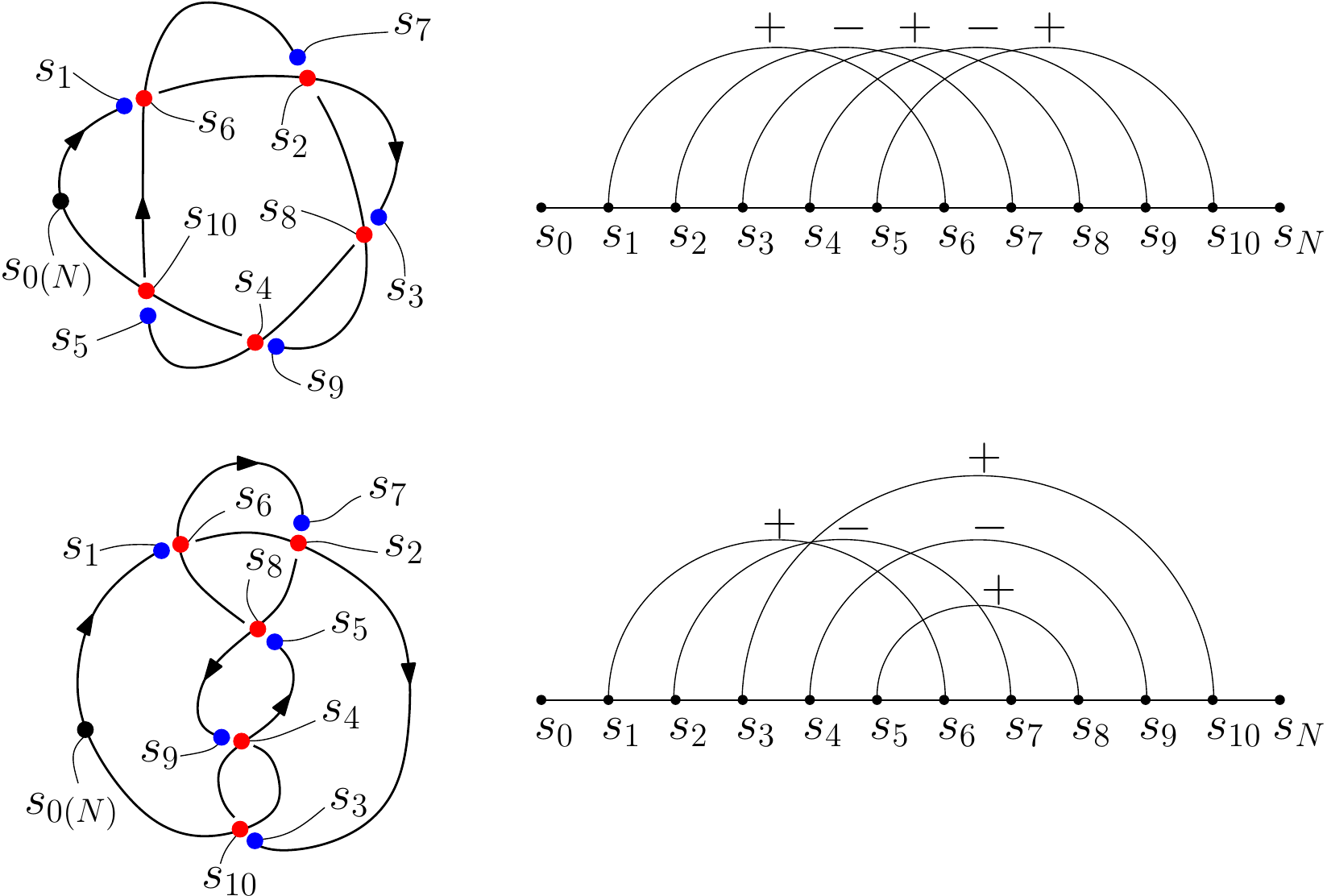}
	\caption{Arc diagrams corresponding to the two prime knots $5_1$ and $5_2$. Distances on the arc diagrams are not drawn to scale.}
	\label{fig:bowdiagexplained}
\end{figure}

Diagrams generated in this manner will henceforth be referred to as arc diagrams. Two examples illustrate these ideas in Figure \ref{fig:bowdiagexplained}. 

\subsection{Boundary conditions on arc diagrams}
\label{sspboundcond}
\begin{figure}[h]
	\centering
		\includegraphics[width=0.4\textwidth]{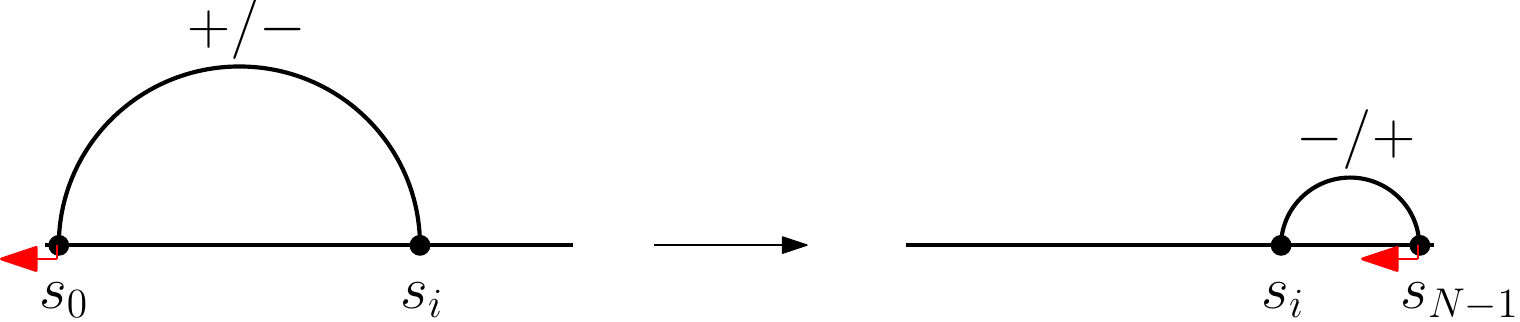}
	\caption{Boundary conditions on arc diagrams: moving an arc foot across the boundary $s_0 =s_N$ results in a sign change for the arc. (This is obviously reversible.)}
	\label{fig:bowboundcond}
\end{figure}

Consider an arc diagram, generated from a knot diagram (i.e., from a closed loop), with an arc between $s_0$ and $s_i$, i.e., one arc foot is on the boundary $s_0 =s_N$. An \textbf{R0} move is executed such that the crossing is now between $s_i$ and $s_{N-1}$. Due to the ordering convention set out above, the sign on the arc diagram would now change, since $s_0<s_i<s_{N-1}$. This is illustrated in Figure \ref{fig:bowboundcond}. Arc diagrams therefore have periodic boundary conditions (due to the periodicity in the arc length coordinate), but a sign change occurs when one arc foot is moved across the boundary $s_0 =s_N$.

\subsection{Reconstructing the knot: arc diagrams and the Gauss code}

Our arc diagram representation of crossings of knots is very similar to the Gauss code. We describe the Gauss code briefly, drawing on the discussions of Kauffman \textit{et al.} \cite{kauffman2006virtual,kauffman1999virtual}.

\begin{figure}[H]
	\centering
		\includegraphics[width=0.5\textwidth]{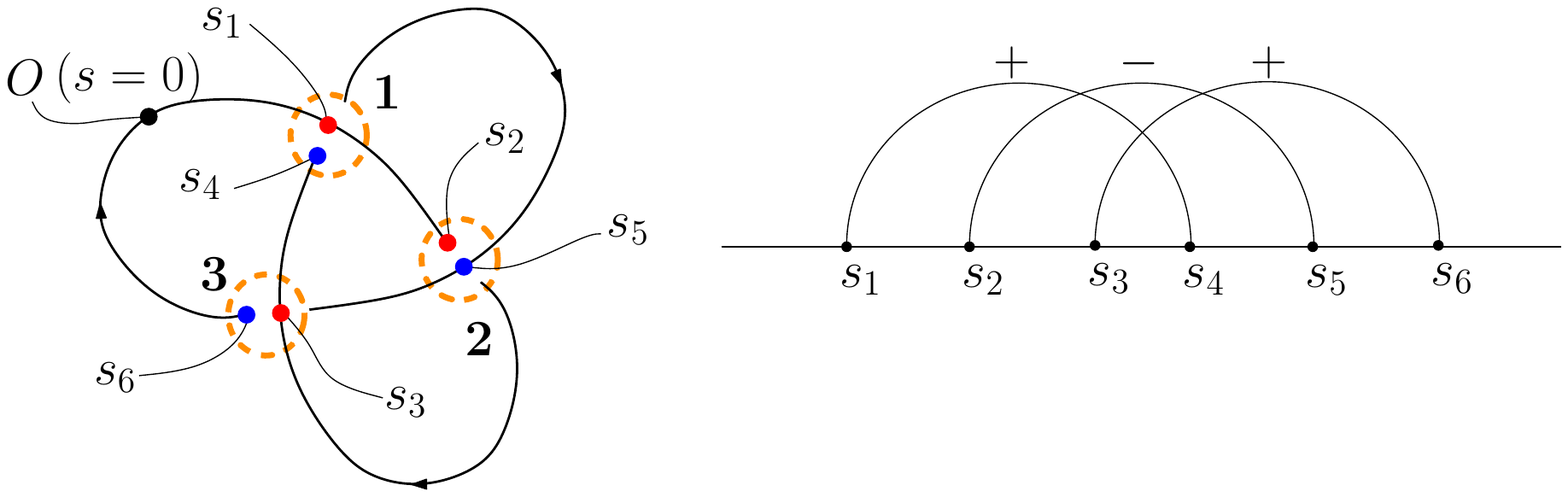}
	\caption{The trefoil knot is shown on the left. An origin and an orientation have been chosen and the three crossings (circled in orange) have been numbered. The $s$ co-ordinate of each strand at the crossings is indicated by $s_i$, $i = 1,2,\ldots,6$ (not to scale). On the right the corresponding arc diagram is shown.}
	\label{fig:trefoil_labelled}
\end{figure}

The Gauss code is a sequence of labels for the crossings of a knot. Each label (crossing) is repeated twice, since each crossing would be encountered twice while walking once along the entire length of any unbroken knot. In addition to the crossing sequence, a Gauss code also records whether a particular strand segment is at the top or bottom of a given crossing. In Figure \ref{fig:trefoil_labelled} we show a trefoil knot and its corresponding arc diagram. The Gauss code corresponding to this trefoil is
\beq
g_{\textrm{tref.}} = O1U2O3U1O2U3,
\label{trefoilgauss}
\eeq
where $O$ and $U$ refer to ``over'' and ``under'', respectively. This sequence contains the same information as the arc diagram in Figure \ref{fig:trefoil_labelled}. Following the $s$ axis of the arc diagram from left to right, we see that the signs of the arcs encountered are analogous to the sequence of $O$s and $U$s. Since each arc corresponds to a crossing, it is trivial to reconstruct the Gauss code in equation\eref{trefoilgauss} from the arc diagram. The arc diagram, however, further records the arc distance between consecutive crossings along the projected arc-length coordinate, and not only their order.

\begin{figure}[H]
	\centering
		\includegraphics[width=0.2\textwidth]{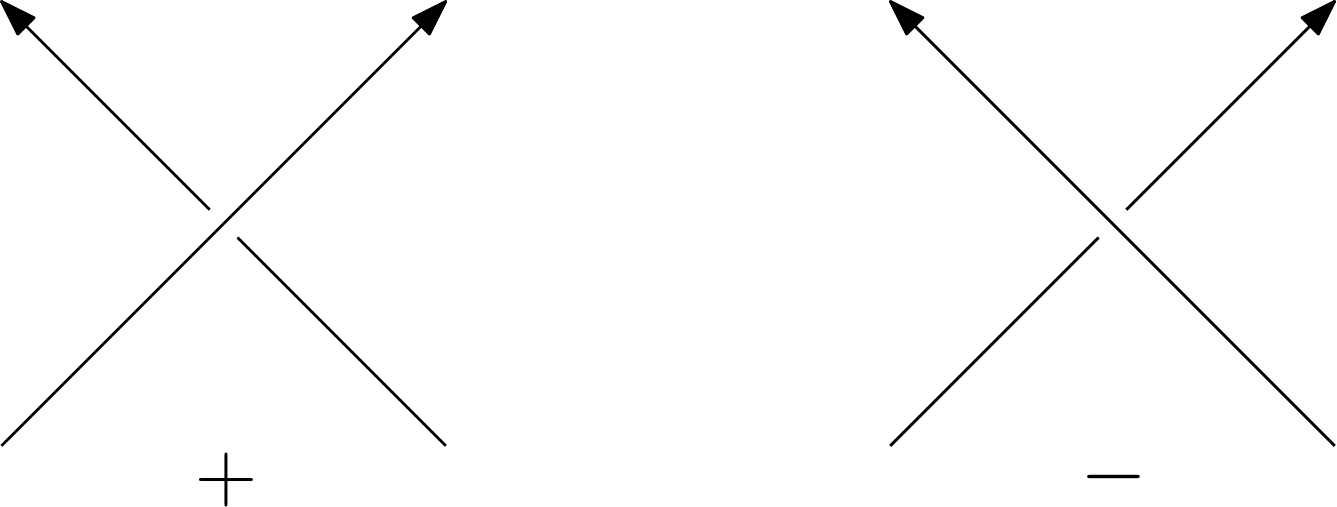}
	\caption{Standard knot theory convention for assigning orientation to a crossing according to its ``handedness''.}
	\label{fig:orientation}
\end{figure}

An extension of the standard Gauss code is the signed Gauss code, which further records the orientation of each crossing, as defined in Figure \ref{fig:orientation}. This orientation is denoted as $+$ if a crossing is ``right-handed'' and as $-$ if it is ``left-handed''. 
The signed Gauss code for the trefoil knot in Figure \ref{fig:trefoil_labelled} is
\beq
g^{(s)}_{\textrm{tref.}} = O1+U2+O3+U1-O2-U3-.
\eeq
Note that this is \textit{not} the same convention chosen to allocate signs to crossings in our arc diagrams. Indeed, the $+$ and $-$ signs of arcs denote the $O$s and $U$s of a Gauss code, i.e., the overpasses and underpasses. Arc diagrams do not capture the orientation of crossings as defined in Figure \ref{fig:orientation}. 

A natural question arises: can a knot diagram be reconstructed unambiguously from an arc diagram? Indeed, there exists an algorithm for reconstructing a knot shadow from a particular Gauss code (this code need not be be signed) \cite{kauffman1999virtual}. The one proviso here is that the Gauss code underlying the construction be planar, i.e., that there is no need to introduce virtual crossings during the reconstruction of the planar shadow \cite{kauffman2006virtual,kauffman1999virtual}. (Knot diagrams with virtual crossings do not have physical realisations as embeddings in three-dimensional space.) Since we are considering classical (read ``non-virtual'') knots, this requirement is trivially satisfied: we work with planar Gauss codes that were generated from a real knot. For such Gauss codes, the reconstruction of the knot shadow is possible up to isotopy on a sphere \cite{goussarov1998}. As regards over- and underpasses, reconstruction using the aforementioned algorithm leaves one arbitrary initial choice in crossing orientation, but the orientation of the remaining crossings is fixed by the order of over- and underpasses \cite{kauffman1999virtual}. Arc diagrams contain the same information as a Gauss code and thus one may reconstruct the knot shadow from a given arc diagram (up to an initial choice of crossing orientation). This is crucial to our discussion, as we are considering the rules that relate topologically equivalent knots. We omit the explicit discussion of Kauffman's reconstruction algorithm since our focus is on the representation of the Reidemeister moves as rules on arc diagrams.

It should be noted that the manipulation of Gauss codes according to the Reidemeister moves has been studied; see, for instance, the appendices of Kauffman's book \cite{kauffman}.

%%%%%%%%%%%%%%%%%%%%%%%%%%%%%%%%%%%%%%%%%%%%%%%%%%%%%%%%%%%%%%%%%%%%%%%%%%%%%%%%%%%%%%%%%%%%%%%%%%%%%%%%%%%%
%%%%%%%%%%%%%%%%%%%%%%%%%%%%%%%%%%%%%%%%%%%%%%%%%%%%%%%%%%%%%%%%%%%%%%%%%%%%%%%%%%%%%%%%%%%%%%%%%%%%%%%%%%%%
%%%%%%%%%%%%%%%%%%%%%%%%%%%%%%%%%%%%%%%%%%%%%%%%%%%%%%%%%%%%%%%%%%%%%%%%%%%%%%%%%%%%%%%%%%%%%%%%%%%%%%%%%%%%
%%%%%%%%%%%%%%%%%%%%%%%%%%%%%%%%%%%%%%%%%%%%%%%%%%%%%%%%%%%%%%%%%%%%%%%%%%%%%%%%%%%%%%%%%%%%%%%%%%%%%%%%%%%%

\section{Representations of the Reidemeister moves on arc diagrams}
\label{bowdiagsect}
In lieu of the arc diagram representation for the crossing structure of a given knot, we now investigate how the Reidemeister moves would look on arc diagrams. Given an arc diagram, how can we generate equivalent arc diagrams? We shall view such manipulations as dynamical rules for the arc feet, governed by the Reidemeister moves since these relate topologically equivalent knots. For an arc diagram of a given knot, the $+$ and $-$ arcs therein are treated as dynamical objects, whose feet ``diffuse'' between sites on the arc diagram (like particles on a lattice) as various crossings ``slide'' around in the planar projection. The latter processes occur in accordance with the topologically trivial moves illustrated in Figure \ref{fig:movestrivial}. The Reidemeister moves from Figure \ref{fig:moveswithlabels} then determine the interaction rules. 

We now consider each Reidemeister move individually, as seen on segments of a knot in Figures \ref{fig:moveswithlabels} and \ref{fig:movestrivial}. Since we only consider the segments of the knot which are close to the crossings involved, the orientations of line segments will be chosen arbitrarily. In practice, these orientations would be determined by the specifics of the remainder of the knot in question. This will prove to be of particular importance for the representations of \textbf{R2} and \textbf{R3}.

\subsection{The move \textbf{R0}}
\label{R0descript}
\begin{figure}[H]
	\centering
		\includegraphics[width=0.4\textwidth]{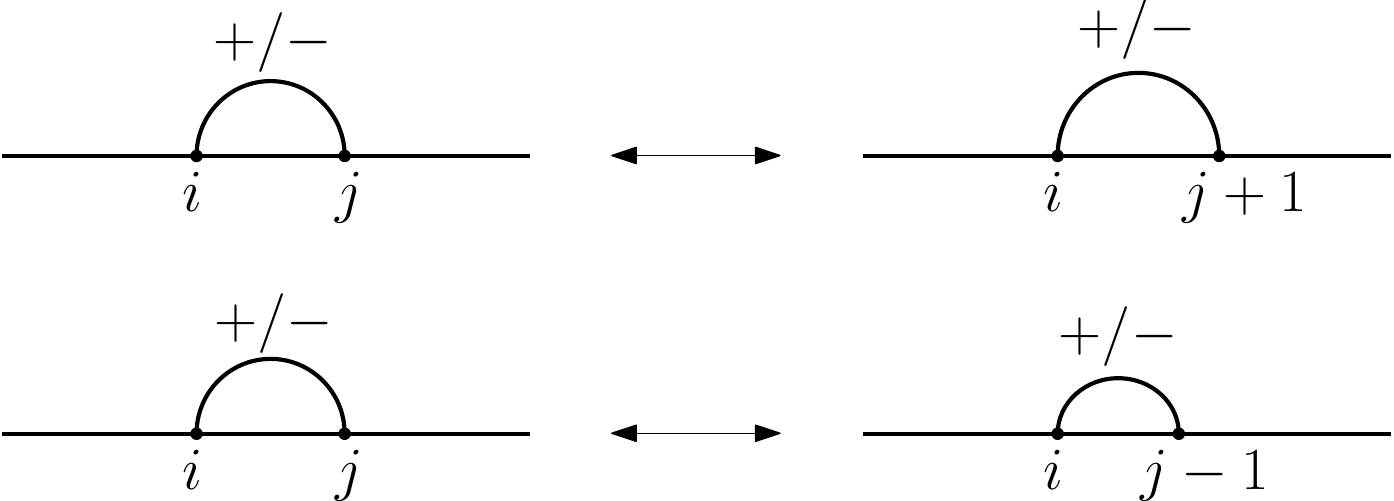}
	\caption{The move \textbf{R0} on an arc diagram. One arc foot ``diffuses'' to an adjacent site on the line, provided that this site is empty. A corresponding scenario where the left foot (labelled $i$) diffuses is not shown here.}
	\label{fig:r0bow}
\end{figure}

Since the first move in Figure \ref{fig:movestrivial} does not alter the distance between crossings, we denote \textbf{R0} the (topologically trivial) move that alters the relative lengths of different segments in the strand; see the second scenario in Figure \ref{fig:movestrivial}. This corresponds to ``diffusion'' of the arc feet on arc diagrams, assumed to be to nearest-neighbouring sites. This move leaves the number of crossings (arcs) unchanged. In our projection we require that at most two strands ever cross each other, and the associated occupation restriction for sites on an arc diagram implies that diffusion can only occur to empty neighbouring target sites. This process is subject to the boundary conditions shown in Figure \ref{fig:bowboundcond}. We illustrate \textbf{R0} on arc diagrams in Figure \ref{fig:r0bow}, where we have suppressed the $s$ label.

\subsection{The move \textbf{R1}}
\label{R1descript}
Consider the forward direction for the first case of the move \textbf{R1} in Figure \ref{fig:moveswithlabels}. As the loop is shortened, the arc feet of the crossing in question would approach each other on the arc diagram. As the loop is removed from the strand, the arc between sites $s_i$ and $s_{i+1}$ is removed from the arc diagram. If instead we consider the reverse direction of the first case of the move \textbf{R1} in Figure \ref{fig:moveswithlabels} where a crossing is created in a strand that was previously crossing-free, the corresponding process on an arc diagram would be the ``creation'' of an arc between nearest neighbouring sites.

\begin{figure}[H]
	\centering
		\includegraphics[width=0.4\textwidth]{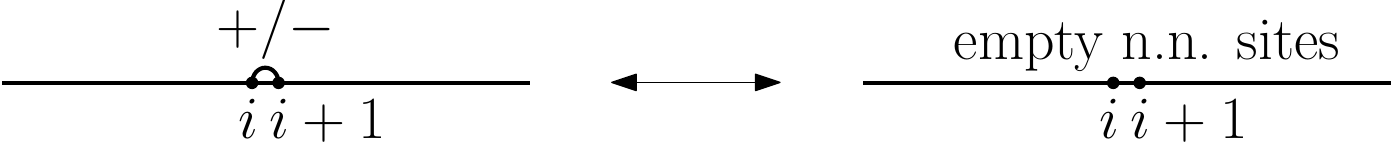}
	\caption{\textbf{R1} on an arc diagram: creation / annihilation of a single arc (of any sign) at neighbouring sites on the line.}
	\label{fig:r1bow}
\end{figure}

The move \textbf{R1} thus involves the creation / annihilation of a single arc at two adjacent sites on the arc diagram. The creation process may only happen if the two sites are unoccupied. This is shown in figure \ref{fig:r1bow}.

Note that the sign of the arc being created / annihilated depends on which of the \textbf{R1} scenarios in Figure \ref{fig:moveswithlabels} is being considered. In principle, a combination move of \textbf{R0} and \textbf{R1} is possible, where an arc that does not have nearest-neighbouring arc feet is inserted to or removed from the arc diagram. We shall assume the scenario where \textbf{R1} only occurs at nearest-neighbouring sites.

\subsection{The move \textbf{R2}}
\label{R2descript}

\begin{figure}[H]
	\centering
		\includegraphics[width=0.4\textwidth]{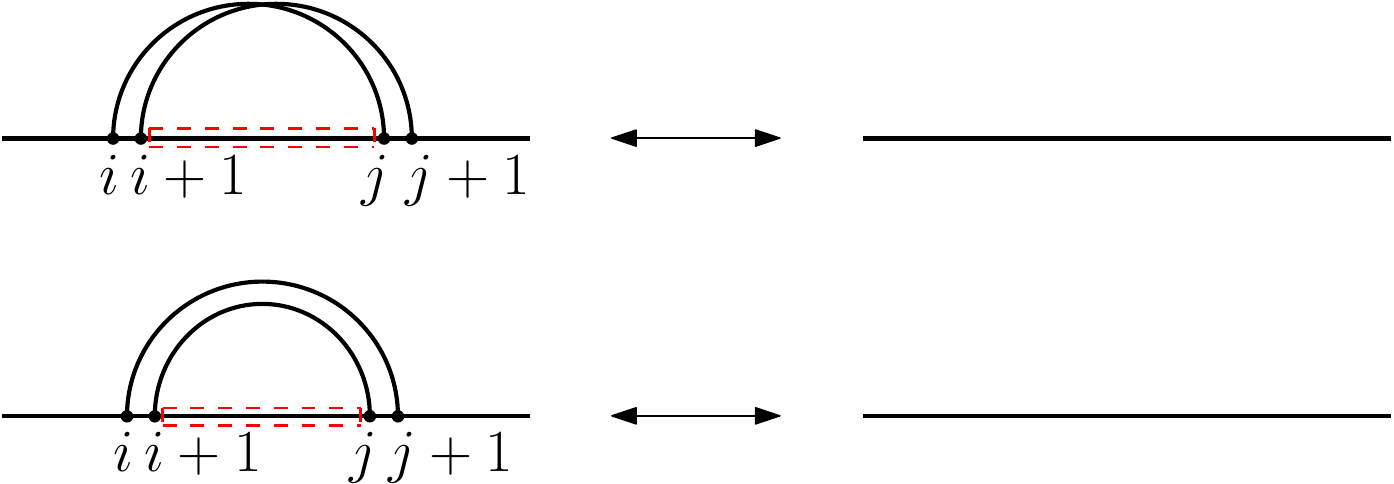}
	\caption{\textbf{R2} on an arc diagram: creation / annihilation of an equal-sign  arc pair. Top: parallel strands, bottom: anti-parallel strands.}
	\label{fig:r2bow}
\end{figure}

Next we consider the move \textbf{R2} in Figure \ref{fig:moveswithlabels}. During this move a pair of crossings is created / annihilated in the knot diagram.  On an arc diagram this is represented by the creation / annihilation of an equal-sign arc pair, where the left feet of both arcs are adjacent and the right feet of both arcs are adjacent. This is illustrated in Figure \ref{fig:r2bow}. For the annihilation process the relative orientation of the two strands is unimportant. Consequently the forward processes shown in Figure \ref{fig:r2bow} can both happen assuming that two arcs \textit{of equal sign} can be found in the correct configuration on a given arc diagram. For the \textbf{R2} annihilation process on an arc diagram it is thus unimportant whether the two arcs cross each other (top of Figure \ref{fig:r2bow}) or whether they are ``nested'' (bottom of Figure \ref{fig:r2bow}).

For the \textbf{R2} creation process (i.e., the reverse processes in Figure \ref{fig:r2bow}), however, care must be taken with the relative orientations of the two strands. Suppose we wish to create an arc pair at points $i$, $i+1$ and $j$, $j+1$ as in Figure \ref{fig:r2bow}. To determine whether the crossings are created on parallel or anti-parallel strands, we must count how many times the strand between $i+1$ and $j$ crosses itself. This corresponds to counting the number of arcs completely contained in the red regions in Figure \ref{fig:r2bow}. For the creation of an \textbf{R2} pair on \textit{parallel strands} this region would contain an odd number of complete arcs. In this case the reverse process at the top of Figure \ref{fig:r2bow} would occur. For the creation of an \textbf{R2} pair on \textit{anti-parallel strands} the red region would contain an even number of complete arcs. In this case the reverse process at the bottom of Figure \ref{fig:r2bow} would occur. (It is not sufficient to simply count the number of arc feet in the red region. If an arc has one arc foot in this region and the other outside, this represents a crossing of the strand between $i+1$ and $j$ with \emph{another} strand in the knot, which does not affect the relative orientation of the strands involved in the \textbf{R2} move.)

\subsection{The move \textbf{R3}}
\label{R3descript}

\begin{figure}[H]
	\centering
		\includegraphics[width=0.1\textwidth]{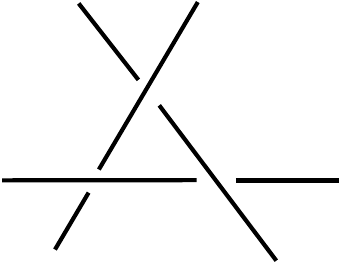}
	\caption{In this configuration it is impossible for one strand to move past the crossing of the two remaining strands.}
	\label{fig:r3fail}
\end{figure}

\begin{figure}[H]
	\centering
		\includegraphics[width=0.4\textwidth]{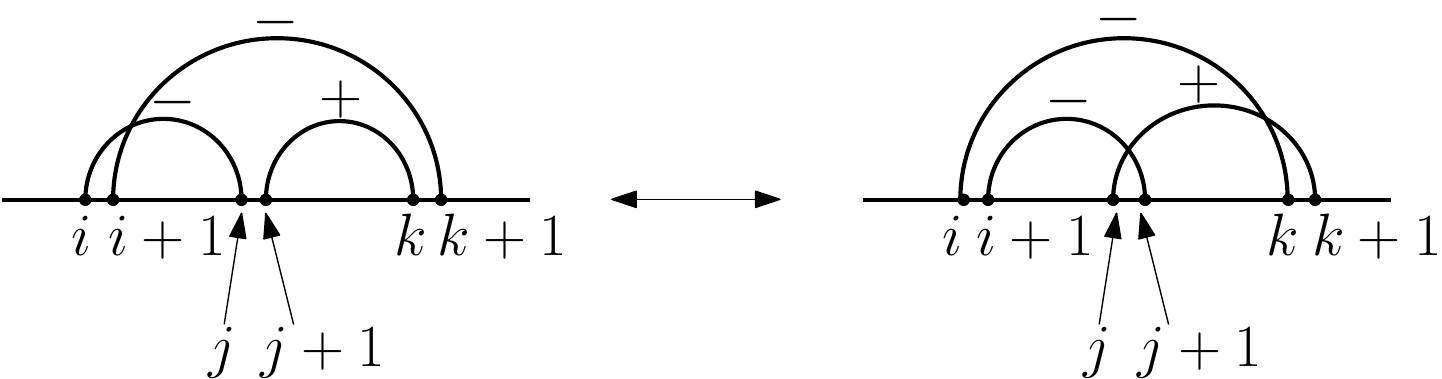}
	\caption{\textbf{R3} on an arc diagram. This particular arrangement of signs corresponds to scenario from Figure \ref{fig:moveswithlabels}. Execution of the move results in exchange of positions of nearest neighbour arc feet.}
	\label{fig:r3bow}
\end{figure}

Finally we take a look at the last Reidemeister move in Figure \ref{fig:moveswithlabels}. Again, the orientations of strands in that figure were chosen arbitrarily and points on the strands were labelled correspondingly. The move is applied to the strands in two parts: the strand containing points $s_j$ and $s_{j+1}$ is moved past the crossing of the other two, and then the other two strands are shifted along to return to a convenient configuration for labelling. (This is technically a combination of \textbf{R0} and \textbf{R3}, but is absorbed into our definition of the latter move.) Note that the move \textbf{R3} can only be executed when one of the three strands has an over- and an undercrossing, and the remaining two strands either have two overcrossings or two undercrossings. In the case where all three strands have an overcrossing and an undercrossing, it would be impossible to move one of the strand past the crossing of the other two; see Figure \ref{fig:r3fail}.
In terms of arc diagrams, the execution of \textbf{R3} translates into three pairwise exchanges of the positions of neighbouring arc feet. An example of this is shown in Figure \ref{fig:r3bow}.
Precluding the scenario in Figure \ref{fig:r3fail} necessitates restrictions on the sign combinations of arcs involved such that two of the three nearest-neighbouring pairs of arc feet have unequal signs, and the third has equal signs.

%%%%%%%%%%%%%%%%%%%%%%%%%%%%%%%%%%%%%%%%%%%%%%%%%%%%%%%%%%%%%%%%%%%%%%%%%%%%%%%%%%%%%%%%%%%%%%%%%%%%%%%%%%%%
%%%%%%%%%%%%%%%%%%%%%%%%%%%%%%%%%%%%%%%%%%%%%%%%%%%%%%%%%%%%%%%%%%%%%%%%%%%%%%%%%%%%%%%%%%%%%%%%%%%%%%%%%%%%
%%%%%%%%%%%%%%%%%%%%%%%%%%%%%%%%%%%%%%%%%%%%%%%%%%%%%%%%%%%%%%%%%%%%%%%%%%%%%%%%%%%%%%%%%%%%%%%%%%%%%%%%%%%%
%%%%%%%%%%%%%%%%%%%%%%%%%%%%%%%%%%%%%%%%%%%%%%%%%%%%%%%%%%%%%%%%%%%%%%%%%%%%%%%%%%%%%%%%%%%%%%%%%%%%%%%%%%%%

\subsection{Summary of allowed ``dynamics'' on arc diagrams}

\bi
\item Only one arc foot is allowed per site on the arc diagram, since at most two strands may cross each other in the projection.
\item Arc feet may ``diffuse''  on the arc diagram. This may only occur if the nearest neighbouring target sites are unoccupied. This process corresponds to relative lengthening / shortening of various strands in the knot diagram, as governed by \textbf{R0}. If two arc feet are in adjacent sites, they may not move past each other. See Section \ref{R0descript} for more details.
\item Arcs on the arc diagram ``interact'' according to the remaining Reidemeister moves:
\bi 
\item \textbf{R1}: a single arc (of either sign) may be created or annihilated at nearest neighbouring sites on the arc diagram. This corresponds to introducing new single loops into the knot projection, or removing single loops. For the creation process, adjacent nearest neighbouring sites on the arc diagram must be unoccupied. See Section \ref{R1descript} for more details.
\item \textbf{R2}: equal sign arc pairs are created or annihilated on the arc diagram. For the creation process, the two pairs of nearest neighbouring target sites must be unoccupied, and relative orientation of the strands must be taken into consideration. This corresponds to moving two strands across each other or separating them on the knot diagram. See Section \ref{R2descript} for details.
\item \textbf{R3}: given that one of the allowed triplet combinations of arcs is present, nearest neighbouring arc feet can exchange sites on the arc diagram. See Section \ref{R3descript} for more details.
\item Importantly, ONLY \textbf{R3} allows arc feet to exchange sites on the arc diagram. \textbf{Without this move, arc feet cannot move past each other}.
\ei
\ei

\subsection{Prime knots and their representation}
\label{primesect}
As mentioned, prime knots are the ``simplest'' knots in that they cannot be reduced to contain fewer crossings through some sequence of Reidemeister moves. A theorem by Schubert \cite{schubert1949} states that any knot may be expressed uniquely as the connected sum of prime knots. (This can be viewed as cutting prime knots open and splicing them together.) In this sense prime knots provide a categorisation scheme for fundamental (i.e., undecomposeable) knots according to the number of crossings they have. Extensive tables of prime knots are readily available \cite{hoste1998}.

Prime knots will be particularly relevant for us. Consequently our labelling schemes must indeed distinguish between different prime knots with the same number of crossings. Our representations of the Reidemeister moves may also not alter the minimal number of crossings for a given prime knot. These requirements are indeed met, as is manifest in the distinct arc diagrams in in Figure \ref{fig:bowdiagexplained}. Furthermore it is easy to verify that none of the Reidemeister moves on arc diagrams (as set out in Section \ref{bowdiagsect}) can reduce the number of arcs (i.e., crossings) for the examples in Figure \ref{fig:bowdiagexplained}. The minimal number of crossings of prime knots is maintained as required, since no crossings can be removed through \textbf{R1} or \textbf{R2}, and no triplet of arcs exists that allows the execution of \textbf{R3}.
%
%; see Figure \ref{fig:8triangles} and Figure \ref{fig:r3bow}
%
%\begin{figure}[H]
	%\centering
		%\includegraphics[width=0.9\textwidth]{knots5152.pdf}
	%\caption{The two prime knots $5_1$ (top) and $5_2$ (bottom) and their associated bow diagrams. These knots both have five crossings, but they are topologically distinct. Distances between bow feet have been rescaled. The corresponding \ssp s are not shown since they may easily be reconstructed from the bow diagrams.}
	%\label{fig:knots5152}
%\end{figure}
%

Indeed, it is an interesting question to ask what is the prime knot underlying some randomly generated knot which is not in its simplest form (i.e., the form with the least number of crossings). In Appendix \ref{simchapter} we shall present some ideas on how the representation of arc diagrams could perhaps be used in the setting of a Monte Carlo type simulation to address such matters.

\subsection{Arc diagrams: what have we achieved?}

With arc diagrams we have a simple graphical representation of the crossing structure of knots, where information about distances between crossings in the projection is retained. We record arcs that each have two positional degrees of freedom (the position of the two arc-feet) and a sign (indicating an over- or undercrossing). This representation has numerous benefits. It allows for quick identification of scenarios where the various Reidemeister moves may be executed. Occupancy restrictions on sites of the arc diagram amount to a simple \textit{one-dimensional} (albeit sometimes non-local) check. The nearest-neighbour checks for the various Reidemeister moves are also easily realised. Consequently the generation of various equivalent arc diagrams and, by implication, knot diagrams is now a simple task.

Topology conservation through the Reidemeister moves was translated into dynamical rules on arc diagrams. The aim is now to encode these rules into stochastic dynamics on a lattice representation of arc diagrams. We present this discussion in the language of master equations for a system with restricted occupation numbers, and establish an operator formalism for the Reidemeister moves. The purpose is to explore how the topological constraints affect dynamical quantities and correlations. Section \ref{doisectionmain} contains an overview of operator techniques for master equations for reaction-diffusion systems. (Readers familiar with these techniques may wish to proceed directly to Section \ref{doisectbow}.) Some relevant applications will be considered. Drawing on these examples, we then derive a formalism for the dynamics on arc diagrams and study densities and correlators in Section \ref{doisectbow}.

\section{Review of some operator techniques for reaction-diffusion systems}
\label{doisectionmain}

Many physical systems can be modelled in terms of processes such as creation, annihilation, diffusion and reactions of particles on a lattice. Occupation numbers for lattice sites provide a natural language to represent states of a reaction-diffusion system. Typically the aforementioned processes are described in terms of stochastic differential equations, such as master equations or Fokker-Planck equations, which govern the time-evolution of the probability that the system occupies each state. Standard references include the books of Gardiner \cite{gardiner} and Risken \cite{risken}. 

In 1976 Doi presented an elegant formalism for translating master equations (describing the rate of change of probability in bosonic occupation number systems) into a field theoretical description \cite{doi1976}. Peliti later expanded on this work in an extensive article \cite{peliti1985} where descriptions of birth and decay processes on a lattice are discussed. This formalism stemming from Doi's work has become a broadly-used tool in the study of reaction-diffusion systems, and has opened the door to powerful field theoretic techniques such as renormalisation techniques \cite{mattisglasser1998, tauber2005}. Typically bosonic systems are considered in this setting, i.e.,  occupation numbers of individual lattice sites or states are not restricted in any way. There exist, however, prescriptions for systems with restricted occupation numbers \cite{sandowtrimper1993,patzlafftrimper1994,bruneljopa,brunelarxiv}. One may expect that the extension to this formalism simply involves replacing bosonic degrees of freedom with fermionic ones on an operator level. This is indeed not the case, since fermions encode an anti-symmetry under exchange of particles which is not desired in this setting: we are considering dynamic processes on lattice sites and simply want the occupation number of each site to be restricted. The solution to this involves the introduction of so-called ``paulions'' which have both fermionic attributes (restricted occupation number) and bosonic attributes (symmetry under exchange of particles). Systems with occupation number restrictions have been studied in the context of aggregation reactions \cite{sandowtrimper1993, rudavets1993} and frameworks have been suggested for finding their classical actions in terms of modified Grassmann variables \cite{patzlafftrimper1994} {and in the context of fermionic field theories \cite{bruneljopa,brunelarxiv}}. It may be of interest that master equations for some one-dimensional non-equilibrium systems can be written as Schr\"odinger equations describing certain quantum chains; see, for instance, \cite{alcaraz1993} and references therein. 

In the following sections we provide a brief outline of Doi's formalism and some basic extensions thereof, referring to some standard examples. We point out in particular how one may arrive at differential equations for the time-evolution of densities and correlators under stochastic dynamics. This mathematical toolbox will later be used when recasting the rules for crossing dynamics into master equations and developing an associated operator formalism.

\subsection{Mapping master equations onto an operator formalism}
\label{doisection1}
Here we present a brief overview of the techniques alluded to above. The discussion is an amalgamate of the early ideas of Doi \cite{doi1976}, and the later extensions thereof by Peliti \cite{peliti1985}, Mattis and Glasser \cite{mattisglasser1998} and T\"auber \etal \cite{tauber2005}. In Section \ref{doisection2} we discuss systems with restricted occupation numbers. Here we begin by considering a bosonic system whose states are labeled completely by a vector of occupation numbers on $N$ lattice sites,
\beq
\mathbf n = (n_1, n_2,\ldots,n_i,\ldots,n_N), \quad n_i = 0,1,2,\ldots.
\label{nvector}
\eeq
The dynamics of such systems may be understood in terms of master equations for the rate of change of the probability for certain configurations,
\beq
\partial_t P(\mathbf n|t) = \sum_{\mathbf n'} \left\{ \omega_{(\mathbf n'\rightarrow\mathbf n)} P(\mathbf n'|t) - \omega_{(\mathbf n\rightarrow\mathbf n')} P(\mathbf n|t) \right\}.
\label{master1}
\eeq
%Probability is conserved, as is seen by summing both sides under the assumption that probabilities are initially normalised. Further, for any choice of non-negative initial probabilities, equation (\ref{master1}) cannot generate negative probabilities since the only negative contribution to a given $P(\mathbf n |t)$ is proportional to $P(\mathbf n |t)$ itself. 
The first term on the right of (\ref{master1}) indicates flux from all states $\mathbf n'$ that could precede the sate $\mathbf n$ (henceforth termed ``precursor states''), and the second indicates flux to all states $\mathbf n'$ that could result from  $\mathbf n$ (henceforth termed ``descendant states''). The transition rates $\omega$ need not be simple constant linear quantities, but typically depend on the configurations $\mathbf n$ and $\mathbf n'$. 
%The condition of detailed balance is met if each term in this summation vanishes at equilibrium, $\omega_{(\mathbf n'\rightarrow\mathbf n)} P(\mathbf n'|t) = \omega_{(\mathbf n\rightarrow\mathbf n')} P(\mathbf n|t).$ 

It is the specific dynamical processes of the system (diffusion, creation, annihilation, aggregation, etc.) that relate the state $\mathbf n$ to precursor and descendant states. Consider diffusion of a particle from site $i$ to site $j$. A possible precursor state to $\mathbf n$ under this process is the state
\beq
\mathbf n' = (n_1, n_2,\ldots,n_i+1,\ldots,n_j-1,\ldots,n_N),
\label{precursor}
\eeq
whereas a possible descendant state from $\mathbf n$ is
\beq
\mathbf n'' = (n_1, n_2,\ldots,n_i-1,\ldots,n_j+1,\ldots,n_N).
\label{descendant}
\eeq

We introduce for each site a set of bosonic creation and annihilation operators obeying
\beq
[a_i,a_j^\dagger] = \delta_{i,j} \quad \textrm{and} \quad [a_i,a_j]=[a_i^\dagger,a_j^\dagger]=0.
\label{bosonoperators}
\eeq
Here $[A,B] = AB - BA$ is the standard commutator. The $N$-site vacuum state,
\beqa
\ket{\mathbf 0} = \ket 0 \otimes\ket 0 \otimes\ldots\otimes \ket 0  \equiv \ket{n_1 = 0,\ldots,n_N = 0},
\label{vac}
\eeqa
vanishes under the action of any $a_i$. A general occupation number state is written as
\beq
\ket{\mathbf n} = \prod_{i=1}^N (a_i^\dagger)^{n_i}\ket{\mathbf 0}=\ket{n_1,\ldots,n_N}.
\label{nstate}
\eeq
Normalisation here differs from that usually considered in a quantum mechanical setting: pre-factors of $\frac 1 {\sqrt{n_i!}}$ have been omitted. This convention has been adopted ubiquitously in the literature pertaining to this method since it simplifies later steps. The states in equation (\ref{nstate}) are orthogonal with respect to the following inner product,
\beq
\overlap{\mb n}{\mb n'} = \prod_{i=1}^N (n_i!)\,\delta_{n_i,n_i'}.
\label{innerprod}
\eeq
The operators in (\ref{bosonoperators}) act on the occupation number state (\ref{nstate}) as follows,
\beq
a_i\ket{\mathbf n} = n_i\ket{n_1,\ldots,n_i-1,\ldots,n_N} \quad \textrm{and} \quad a_i^\dagger\ket{\mathbf n} = \ket{n_1,\ldots,n_i+1,\ldots,n_N},
\label{actiononn}
\eeq
where departure from standard normalisation should again be noted. 

To connect the operator formalism and the master equation (\ref{master1}) we define a state vector
\beq
\ket{\phi(t)} \equiv \sum_{\mathbf n} P(\mathbf n|t) \ket{\mathbf n},
\label{phit}
\eeq
which is just a sum over all occupation states of the system, appropriately weighted by their respective (time-dependent) probabilities. The probability of a particular state $\mathbf{\tilde n}$ may be recovered through
\beq
P(\mb{\tilde n}|t) = \frac{1}{\prod_i \tilde{n}_i!}\overlap{\mb{\tilde n}}{\phi(t)}.
\label{probfromphi}
\eeq
This equation allows the identification of a Liouvillian operator $\hat L$ that governs the time-evolution of the state vector,
\beq
\partial_t \ket{\phi(t)} = \hat L \ket{\phi(t)}.
\label{timeevol}
\eeq
$\hat L$ encodes the physical processes that relate current, precursor and descendant states in the master equation (\ref{master1}). In Section \ref{doiexamples} we shall discuss some basic examples of applications; also see \cite{mattisglasser1998, tauber2005} for discussions of pair annihilation / creation, diffusion, aggregation, multi-species processes and higher-order decay processes. 

Note that $\hat L$ need not be Hermitian in general. Indeed, probability conservation is not encoded through unitary time-evolution as would be the case in a quantum mechanical setting. Instead we define a ``sum'' state
\beq
\ket{s} \equiv e^{\sum_{i=1}^N a_i^\dagger}\ket{\mb 0} = \sum_{\mb n} \ket{\mb n}.
\label{projstate}
\eeq
This state is a uniform superposition of all possible combinations of occupation numbers, as can be seen by Taylor expansion of the exponential, subject to the commutation relations in (\ref{bosonoperators}). \Eref{projstate} is an eigenstate of all $a_i$ with an eigenvalue of unity. Consequently, due to the orthogonality condition (\ref{innerprod}), $\overlap{s}{\mb n} = 1$ for all states $\ket{\mb n}$. Supposing $\hat L$ is known, the formal solution to (\ref{timeevol}) is
\beq
\ket{\phi(t)} = e^{\hat L t} \ket{\phi(0)}
\label{phit2}
\eeq
for some initial state $\ket{\phi(0)}$. For any initial state it must hold that $1 = \overlap{s}{\phi(t)} = \matrel{s}{e^{\hat L t} }{\phi(0)}$. Therefore probability conservation implies the requirement
\beq
\bra{s} \hat L = 0.
\label{probcons}
\eeq
This is automatically satisfied by Liouvillians derived from probability-conserving master equation \cite{tauber2005}. The condition (\ref{probcons}) is thus the analogue of hermiticity in quantum mechanics. For observable quantities $A(\mb n,t)$ that depend on the occupation numbers, time-averages may be calculated through
\beqa
\langle A(t) \rangle &=& \sum_{\mathbf n} P(\mb n|t) A(\mb n) = \sum_{\mathbf n} P(\mb n|t) \matrel{s}{\hat A}{\mb n} = \matrel{s}{\hat A}{\phi(t)}.
\label{timeave}
\eeqa
The operator $\hat A$ is obtained by the association $n_i \leftrightarrow \hat n_i \equiv a_i^\dagger a_i$. The average $\langle \cdot \rangle$ here refers to an average over all realisations of the stochastic dynamics encoded in the master equation. From (\ref{timeevol}) and (\ref{probcons}) follows the equation of motion for observables,
\beqa
\partial_t \langle A(t) \rangle = \matrel{s}{\hat A \hat L}{\phi(t)} = \matrel{s}{[\hat A, \hat L]}{\phi(t)}.
\label{dttimeave}
\eeqa
This allows us to calculate the rate of change of averages and correlators of various quantities \cite{sandowtrimper1993,patzlafftrimper1994}. In particular we point out for the single-site occupancy ($\hat n_I$) that
\beq
\partial_t \langle \hat n_I \rangle (t) = \matrel{s}{[a_i^\dagger a_i, \hat L]}{\phi(t)}.
\label{singlesiteave}
\eeq
This can, for instance, be used to study time-evolution of the two-site correlator
\beq
C_{I,J} = \langle n_I n_J \rangle - \langle n_I \rangle \langle n_J \rangle.
\label{correlator}
\eeq

This establishes a mapping between master equations and a time-evolution equation formulated in terms of creation and annihilation operators. Average quantities can be calculated through (\ref{dttimeave}), and the operator formulation opens the door to techniques such as time-slicing to obtain a field theoretical representation of equation (\ref{phit2}) \cite{peliti1985,mattisglasser1998,tauber2005,patzlafftrimper1994}. This has been addressed, for instance, through renormalisation methods \cite{tauber2005}. It is further possible to calculate the corresponding action through a coherent state path integral; a discussion for bosonic systems is found in \cite{tauber2005}. 

In the present work we wish to express the Reidemeister moves as represented on arc diagrams (see Section \ref{bowdiagsect}) in terms of Liouvillians that are obtained from the corresponding master equations. As stated, the occupation numbers in arc diagrams are subject to certain restrictions that will need to be encoded into this description. To this end the bosonic commutation relations (\ref{bosonoperators}) will need to be modified. We briefly discuss these modifications in a general setting in the following section.

\subsection{Operator formalism for restricted occupation numbers: the paulionic case}
\label{doisection2}
Suppose we repeat the preceding analysis, but for systems with restricted occupation numbers $\mb n = (n_1,\ldots,n_N)$ such that $n_i \in \;\{0,1\},\;\;i = 1,\ldots,N$. We summarise briefly the consequences of this restriction, following the discussions in \cite{mattisglasser1998,tauber2005,sandowtrimper1993}. 

A straightforward replacement of the bosonic commutation relations (\ref{bosonoperators}) with fermionic ones brings about anti-symmetry of states. We merely seek the restriction of occupation numbers, but wish to retain symmetry under particle exchange. Consequently we define operators that commute at different lattice sites but obey fermionic anti-commutation relations on-site, referred to as paulions \cite{mattisglasser1998} since they may be represented in terms of a set Pauli matrices that commute off-site. Explicitly we require
\beqa
\{a_i,a_j^\dagger\} &=& \delta_{i,j}, \nonumber \\
\,[a_i,a_j] &=& [a_i^\dagger,a_j^\dagger] = 0, \nl
a_i^2 &=& (a_i^\dagger)^2 = 0, \nl
\,[a_i, a_j^\dagger] &=& \delta_{i,j}(1-2 a_i^\dagger a_i).
\label{paulions}
\eeqa
Here $\{A,B\} = AB +BA$ is the standard anti-commutator. The last line of equation (\ref{paulions}) is easily seen to be a consequence of the first three. These mixed commutation relations preclude the undesired anti-symmetry of completely anti-commuting fermionic operators. (Paulions may be related to standard fermions through the Jordan-Wigner transformation \cite{mattisglasser1998,bruneljopa,brunelarxiv}, although we shall not use this fact explicitly.)

Barring the modifications in (\ref{paulions}), the remainder of the formalism set out in Section \ref{doisection1} is not altered. States are still labelled by an occupation number vector $\ket{\mb n}$ as in equation (\ref{nstate}), except that $n_i \in \;\{0,1\}$. The sum state $\ket{s}$  from equation (\ref{projstate}) is unchanged. However, the paulionic relations (\ref{paulions}) imply that $\ket s$ is no longer an infinite superposition, but rather a superposition (of $2^N$ terms for $N$ lattice sites) with all possible combinations $n_i \in \;\{0,1\}$.

The definition of the state vector (\ref{phit}), the time-evolution equation (\ref{timeevol}) and the calculation of stochastic averages in equations (\ref{timeave}) and (\ref{dttimeave}) are all unchanged. Any expansions of exponentials of the operators from (\ref{paulions}) will truncate after the first order term, since all paulionic operators square to zero. Conveniently $n_i \in \;\{0,1\}$ further implies that all factorials in the bosonic inner product (\ref{innerprod}) need not be considered explicitly for paulions; this also holds for the mapping from the state vector to the probabilities in equation (\ref{probfromphi}).
The specific dynamics of such a system (and the corresponding Liouvillian) would have to include the restriction of occupation numbers; see, for instance, \cite{sandowtrimper1993}. We consider a few examples of dynamical processes to illustrate the differences between bosonic and paulionic systems in this setting.

\subsection{Examples of applications to dynamical processes}
Here we briefly illustrate the concepts of Section \ref{doisection1} through applications to specific physical processes. In particular, we point out the differences between the bosonic and paulionic descriptions. Throughout we shall assume that we are dealing with a single species (described in terms of operators $a_i$ and $a_i^\dagger$) on a lattice with $N$ sites. The following ideas will be developed for the application to crossing dynamics later.
\label{doiexamples}
\subsubsection{Diffusion}
\label{doidiffusion}
Diffusion on a lattice may be viewed as hopping of particles between nearest-neighbouring sites $i$ and $j$. We base the following discussion loosely on references \cite{tauber2005} and \cite{patzlafftrimper1994}.

\paragraph{Bosonic case} 

For a bosonic system, the master equation for diffusion from site $i$ to site $j$ is
\beq
\partial _t P(\mathbf n | t) = D \sum_{<i,j>} \left\{ (n_i + 1) P(\ldots,n_i+1,n_j-1,\ldots|t) - n_i P(\mathbf n|t) \right\}.
\label{diff1}
\eeq
The summation is over nearest-neighbour sites, and could also be written as $\sum_i \sum_{j(i)} $ where $j(i)$ are nearest neighbouring sites of $i$. $D$ is a diffusion constant. The first term on the right captures the possible precursor states of the state $\ket {\mathbf n}$ (see (\ref{precursor})); the $(n_i+1)$ encodes multiplicity of the process. The second term on the right indicates probability flux \emph{out of} state $\ket {\mathbf n}$ through this process, which can happen in $n_i$ ways. For bosons there is no restriction on the descendant state.
From equation (\ref{probfromphi}), the Liouvillian corresponding to the bosonic diffusion master equation (\ref{diff1}) is found as
\beq
\hat L_{\text{diff., bos.}} = D \sum_{<i,j>} \{ a_j^\dagger a_i - a_i^\dagger a_i \},
\label{Ldiffbos}
\eeq
where the $a$s obey the bosonic relations (\ref{bosonoperators}). It is the \textit{backward} action of $\hat L$ and the factorials in (\ref{probfromphi}) that generate the appropriate prefactors in equation (\ref{diff1}). It is easy to verify that this Liouvillian is probability conserving through equation (\ref{probcons}) since
\beq
\bra{s} a_j^\dagger a_i = \bra{s} a_i = \bra{s} a_i^\dagger a_i.
\label{nons}
\eeq
This follows since $\bra{s} $ is an infinite superposition of all (bosonic) occupation numbers which is unaffected by laddering down once at any site. We can now calculate the average occupation number at site $I$ through equation (\ref{dttimeave}),
\beqa
\partial _t \langle \hat n _I \rangle = \matrel{s}{[\hat n _I, \hat L_{\text{diff., bos.}}]}{\phi(t)} 
= D\sum_{i(I)}\matrel{s}{a_I^\dagger a_i - a_i^\dagger a_I}{\phi(t)} 
%= D\sum_{i(I)}\matrel{s}{a_i^\dagger a_i - a_I^\dagger a_I}{\phi(t)} 
= D\sum_{i(I)} \left\{ \langle \hat n _i \rangle - \langle \hat n _I \rangle \right\}.
\label{dtnIbos}
\eeqa
Summation is over sites $i$ that are nearest neighbours to $I$. The result is a standard discrete diffusion equation. 
%where a positive contribution to $\langle \hat n _I \rangle$ can only occur if neighbouring sites $i(I)$ are occupied, and a negative contribution to $\langle \hat n _I \rangle$ can only occur if there are particles present at site $I$.

A similar calculation may be performed for the two-site correlator (\ref{correlator}); see \cite{patzlafftrimper1994}.
%\beqa
%\partial _t C_{I,J}(t) &=& D \sum_{k(I)}(C_{I,k} - C_{I,J}) + D \sum_{k(J)} (C_{k,J} - C_{I,J}) \nl
%&& -D \;\delta_{<I,J>}\;(\langle n_I \rangle + \langle n_J \rangle-2\langle n _I n_J \rangle) \nl
%&&+D \; \delta_{I,J}\sum_{k(I)}(\langle n_I \rangle + \langle n_k \rangle-2\langle n _I n_k \rangle).
%\eeqa
%Here $\delta_{<I,J>}$ is the Kronecker delta function for nearest-neighbouring sites $I$ and $J$.

%\vspace{0.5cm}

\paragraph{Paulionic case}

For the paulionic case the diffusion master equation reads
\beq
\partial _t P(\mathbf n | t) = D \sum_{<i,j>} \left\{ (n_i + 1) P(\ldots,n_i+1,n_j-1,\ldots|t) - (1-n_j)n_i P(\mathbf n|t) \right\}.
\eeq
In contrast to the bosonic case (\ref{diff1}), the descendant state undergoes a selection. A pre-factor $(1-n_j)$ enforces that diffusion from $i$ to $j$ can only occur \textit{if site $j$ is unoccupied}. Correspondingly, 
\beq
\hat L_{\text{diff., paul.}} = D \sum_{<i,j>} \{ a_j^\dagger a_i - a_j a_j^\dagger a_i^\dagger a_i \},
\label{Ldiffpaul}
\eeq
where the $a$s now obey the paulionic commutation relations (\ref{paulions}) and the operator $a_j a_j^\dagger=1-\hat n_j$ is included with the second term generates. The selection of descendant states in the master equation is thus modified through the occupation number restriction. Other pre-factors in the master equation are unchanged, but now subject to the constraint $n_i \in \{0,1\}$ $\forall i$.

Formally, average quantities may be computed as for the bosonic case but while using the paulionic relations (\ref{paulions}) instead of the bosonic ones (\ref{bosonoperators}). Therefore the calculation of commutators with the Liouvillian as in see (\ref{dttimeave}) requires more care. Furthermore, for paulions it is no longer true that $\bra{s} a^\dagger_k  = \bra{s}$ $\forall k$. Instead, for paulions
\beq
\bra{s} a^\dagger_k = \bra{s} a_k a^\dagger_k \quad \textrm{and} \quad \bra{s} a_k = \bra{s} a^\dagger_k a_k.
\label{updown}
\eeq
The paulionic diffusion Liouvillian (\ref{Ldiffpaul}) is easily shown to be probability conserving, using
%\beqa
%\bra{s} \hat L_{\text{diff., paul.}} &=& D \sum_{<i,j>} \bra{s}  \{ a_j^\dagger a_i - a_j a_j^\dagger a_i^\dagger a_i \} \nl
%&=& D \sum_{<i,j>} \bra{s} \{ a_j^\dagger a_i - a_j^\dagger  a_i \} \nl
%&=& 0,
%\eeqa
%where we have used 
(\ref{updown}). Despite the departures from the bosonic case, the density at site $I$ is unchanged.
%\beqa
%\partial _t \langle \hat n _I \rangle &=& \matrel{s}{[\hat n _I, \hat L_{\text{diff., paul.}}]}{\phi(t)} \nl
%&=& D\sum_{i(I)}\matrel{s}{a_I^\dagger a_i - a_i^\dagger a_I}{\phi(t)} \nl
%&=& D\sum_{i(I)}\matrel{s}{a_I a_I^\dagger a_i^\dagger a_i - a_i a_i^\dagger a_I^\dagger a_I}{\phi(t)} \nl
%&=& D\sum_{i(I)} \left\{ \langle \hat n _i \rangle - \langle \hat n _I \rangle \right\},
%\label{dtnIpaul}
%\eeqa
%where we first used (\ref{updown}) and then identified $a_i a_i^\dagger = 1 - a_i^\dagger a_i = 1- \hat n_i$. 
Indeed, for a non-interacting diffusion system the differences between restricted and unrestricted occupation numbers is only seen on the level of correlators \cite{patzlafftrimper1994}.

\subsubsection{Particle creation and annihilation}
\label{doicran}
Again we contrast the bosonic and paulionic cases.

\paragraph{Bosonic case}

The master equation for creation of a boson at site $i$ is
\beq
\partial _t P(\mathbf n | t) = g \sum_{i} \left\{ P(\ldots,n_i-1,\ldots|t) - P(\mathbf n|t) \right\},
\label{mastercrbos}
\eeq
where $g$ determines the rate of creation. The precursor state must have one less particle at that site, and the bosonic system can exit the current state through creation of a particle at site $i$ without any restrictions. These conditions are encoded in the following Liouvillian,
\beq
\hat L _{\textrm{cr.,bos.}} = g\sum_{i=1}^N \left(a_i^\dagger - 1  \right),
\label{Lcrbos}
\eeq
as found in \cite{tauber2005}. This mapping is easily established through (\ref{actiononn}) and (\ref{probfromphi}).
For boson annihilation at site $i$ the master equation is
\beq
\partial _t P(\mathbf n | t) = h \sum_{i} \left\{ (n_i+1) P(\ldots,n_i+1,\ldots|t) - n_i P(\mathbf n|t) \right\},
\label{masteranbos}
\eeq
where $h$ is a rate constant. The precursor state to annihilating a particle at $i$ must have one more particle at that site, andthe prefactor $n_i+1$ indicates multiplicity. The system can only exit the current state through annihilation at site $i$ if there is indeed a particle at this site; this can happen in $n_i$ ways. Again we find the corresponding Liouvillian through (\ref{actiononn}) and (\ref{probfromphi}),
\beq
\hat L _{\textrm{an.,bos.}} = h\sum_{i=1}^N \left(a_i -  a_i^\dagger a_i \right).
\label{Lanbos}
\eeq

\paragraph{Paulionic case}

In the paulionic case we must consider how occupation number restrictions affect the Liouvillians. We present here some extensions to the results of \cite{tauber2005}. If $n_i \in \{0,1\}$ $\forall i$, a particle can only be created at site $i$ if it is unoccupied. The corresponding paulionic Liouvillian is
\beq
\hat L _{\textrm{cr.,paul.}} = g\sum_{i=1}^N \left(a_i^\dagger - a_i a_i^\dagger  \right).
\label{Lcrpaul}
\eeq
The first term in the summation is as with (\ref{Lcrbos}). Since $(a_i^\dagger)^2 = 0$, the precursor state must indeed be unoccupied at $i$, as required. The second term, however, is altered: a paulion can only be created at site $i$ if this site is unoccupied. This is ensured by the operator $a_i a_i^\dagger = 1-a_i^\dagger a_i $.

For annihilation to occur at site $i$, the only requirement is that some particle must be at this site. Consequently the paulionic Liouvillian for this process is the same as the bosonic operator in (\ref{Lanbos}),
\beq
\hat L _{\textrm{an.,paul.}} = h\sum_{i=1}^N \left(a_i -  a_i^\dagger a_i \right).
\label{Lanpaul}
\eeq
Here the occupation number restriction is enforced by $a_i^2 = 0$; again, see (\ref{paulions}). The second term again indicates that the system can only depart from the current state through annihilation of a particle at $i$ if indeed there is a particle at this site; this is enforced by the number operator $a_i^\dagger a_i$ which has eigenvalues of $0$ or $1$ for paulions.

\subsubsection{Other processes and multiple species}

For our purposes the illustrative examples of diffusion and creation / annihilation suffice. The operator formalism can however be applied to a various other processes, including aggregation ($A + A \rightarrow A$) and multi-particle birth and decay processes ($0 \leftrightarrow A^m$, $m\geq 2$). For detailed discussions, see \cite{mattisglasser1998,tauber2005} and the references therein.

%\subsubsection{Multiple species}
\label{doimultispecies}
It is also possible to generalise the above discussions to multi-species processes such as $A+B\rightarrow 0$ or $A+B\rightarrow C$, and ``harvesting''-type reactions ($A + B \rightarrow A$) relevant to aggregation-limited diffusion \cite{sandowtrimper1993}. Extension of the framework from Sections \ref{doisection1} and \ref{doisection2} for two species, $A$ and $B$, simply requires independent creation and annihilation operators $a_i$, $a_i^\dagger$, $b_i$ and $b_i^\dagger$. All operators for species $A$ commute with all operators for species $B$, but operators for each species individually obey the on-site bosonic or paulionic commutation relations, as in (\ref{bosonoperators}) and (\ref{paulions}), respectively. Corresponding Liouvillians can then be defined in terms of these operators.

%%%%%%%%%%%%%%%%%%%%%%%%%%%%%%%%%%%%%%%%%%%%%%%%%%%%%%%%%%%%%%%%%%%%%%%%%%%%%%%%%%%%%%%%%%%%%%%%%%%%%%%%%%%%
%%%%%%%%%%%%%%%%%%%%%%%%%%%%%%%%%%%%%%%%%%%%%%%%%%%%%%%%%%%%%%%%%%%%%%%%%%%%%%%%%%%%%%%%%%%%%%%%%%%%%%%%%%%%
%%%%%%%%%%%%%%%%%%%%%%%%%%%%%%%%%%%%%%%%%%%%%%%%%%%%%%%%%%%%%%%%%%%%%%%%%%%%%%%%%%%%%%%%%%%%%%%%%%%%%%%%%%%%
%%%%%%%%%%%%%%%%%%%%%%%%%%%%%%%%%%%%%%%%%%%%%%%%%%%%%%%%%%%%%%%%%%%%%%%%%%%%%%%%%%%%%%%%%%%%%%%%%%%%%%%%%%%%

\section{Reidemeister moves as stochastic dynamics: Occupation numbers, master equations and Liouvillians for arc diagrams}
\label{doisectbow}
In Section \ref{bowdiagsect} we set out how the crossings of a particular knot may be captured in an arc diagram, and derived rules on these diagrams that encode topology conservation through the Reidemeister moves. We now discuss how an occupation number labelling scheme for arc diagrams, and cast the dynamical rules into stochastic dynamics through operator techniques from Section \ref{doisectionmain}. {As mentioned, the aim is to develop new approaches to studying the role of topology conservation in dynamics of knot projections. The methods considered here have found application in the study of dynamical regimes in systems with birth and decay processes, e.g. \cite{houch,paessensschuetz,baumann}. We represent here the topology conserving manipulations (Reidemeister moves) on knot diagrams as processes in a stochastic dynamical formalism. Although our dynamics are defined on projections (as opposed to three-dimensional knots), we hope that this work is a first step and a new approach to addressing this important question.}

We represent a given arc diagram in terms of a set of occupation numbers,
\beq
\Lambda = \{N_{i,j,\sigma}\},
\label{lambda}
\eeq
where 
\begin{itemize}
	\item $i,j= 1,2,\ldots , N$ label two positions on the line of an arc diagram that has been discretised into $N$ sites,
	\item $N_{i,j,\sigma} = 1$ if there exists an arc of species $\sigma\in \{+,-\}$ between sites $i$ and $j$, and
	\item $i\neq j$ and $i < j$ are assumed (ordering convention).
\end{itemize}
$\Lambda$ is the two-index analogue of the vector of occupation numbers $\mb n$ from (\ref{nvector}). The restrictions on occupation numbers for arc diagrams, as discussed in Section \ref{bowdiagsect}, imply that, for some $i<j$,
\beqa
&& N_{i,j,\sigma} \in \{0, 1\}, \nl
&& N_{i,j,\sigma}=1 \implies N_{i,k,\sigma'}=0 \,\,\forall \,\,k \neq j, \,\,\forall \sigma', \quad \textrm{and} \nl
&& \sum_{k(<i)} \sum_\sigma N_{k,i,\sigma}+\sum_{k(>i)} \sum_\sigma N_{i,k,\sigma} \leq 1.
\label{nconditions}
\eeqa
(In the second line it is implied that the same conditions hold on $N_{k,i,\sigma'}$ if $k<i$.) The aim now is to formulate a master equation for a configuration $\Lambda$,
\beq
\partial_t P(\Lambda|t) = \sum_{\Lambda'}\omega_{(\Lambda'\rightarrow\Lambda)} P(\Lambda'|t) - \sum_{\Lambda''}\omega_{(\Lambda\rightarrow\Lambda'')} P(\Lambda|t).
\label{mebasic}
\eeq
In this case, precursor and descendant states are related to $\Lambda$ through the Reidemeister moves on arc diagrams. 

In the next section we construct occupation number states corresponding to (\ref{lambda}) and consider how the restriction on occupation numbers may be encoded through appropriate operator relations. This is then applied in finding Liouvillians to represent the Reidemeister moves.

\subsection{Operator representation: arc diagrams as occupation number states}
The occupation number restrictions for arc diagrams, as in (\ref{nconditions}), necessitate two species of paulions (see Section \ref{doimultispecies}). The host space for states is a tensor product of the two state spaces,
\beq
\mathcal H = \mathcal S \otimes \mathcal B.
\label{hostspace}
\eeq
Here $\mathcal S$ represents arc feet at \textit{single sites on the line of the arc diagram}. Basis states for $\mathcal S$, $\ket{\mathbf n_{s}}$, are labelled by a vector $\mathbf n_s = (n_1,n_2\ldots,n_N)$ containing individual occupation occupation numbers for sites, as with the vector (\ref{nvector}). Since $n_i \in \{0,1\}$ in arc diagrams, dim$(\mathcal S) = 2^N$. The space $\mathcal B$ represents \textit{arcs between sites}, and has basis states $\ket{\mathbf n_b}$ labelled by the numbers $\{n_{i,j,\sigma}\}$ ($i,j = 1,\ldots,N$ and $\sigma = \pm 1$), indicating the presence of arcs between sites and their species. It is implied that $i<j$ for a given $n_{i,j,\sigma}$. Since there are two species $\sigma \in \{+,-\}$ and we choose two of $N$ sites for each arc, dim$(\mathcal B)= 2 \binom N 2 = N(N-1)$. 

Basis states for $\mathcal S$ and $\mathcal B$ are obtained from the vacuua of the spaces $\mathcal S$ and $\mathcal B$ with paulionic creation operators $a^\dagger_i$ and $b_{i,j,\sigma} ^\dagger$ akin to those in equation (\ref{paulions}), such that
\beq
\ket{\mathbf n_{s}} = \prod_i (a_i^\dagger)^{n_i} \ket{\mb 0} _{\mathcal S}
\label{nsstates}
\eeq
and
\beq
\ket{\mathbf n_{b}}  = \prod_{i,j,\sigma} (b_{i,j,\sigma}^\dagger) ^{n_{i,j,\sigma}} \ket{\mb 0}_{\mathcal B}.
\label{nbstates}
\eeq
The states in equation (\ref{nsstates}) are orthogonal with respect to the usual inner product (\ref{innerprod}), subject to $n_i \in \{0,1\}$. The states in equation (\ref{nbstates}) obey
\beq
\overlap{\mb n_b}{\mb n_b'} = \prod_{i,j=1}^N \prod_{\sigma=\pm}\;\delta_{n_{i,j,\sigma},n_{i,j,\sigma}'}.
\label{innerprod2}
\eeq
Creation and annihilation operators obey paulionic commutation relations, i.e., if \emph{all} indices on two operators are equal they anti-commute, otherwise they commute. For the operators on $\mathcal S$ this is simply given by (\ref{paulions}). For the operators on $\mathcal B$ we have
\beqa
\{b_{i,j,\sigma},b_{i',j',\sigma '}^\dagger\} &=& \delta_{i,i'}\,\delta_{j,j'}\,\delta_{\sigma,\sigma'}, \nonumber \\
\,[b_{i,j,\sigma},b_{i',j',\sigma '}] &=& [b_{i,j,\sigma}^\dagger,b_{i',j',\sigma '}^\dagger] = 0, \nl
(b_{i,j,\sigma})^2 &=& (b_{i,j,\sigma}^\dagger)^2 = 0, \nl
\,[b_{i,j,\sigma},b_{i',j',\sigma '}^\dagger] &=& \delta_{i,i'}\,\delta_{j,j'}\,\delta_{\sigma,\sigma'}\,(1-2 b_{i,j,\sigma}^\dagger b_{i,j,\sigma}).
\label{bowpaulions}
\eeqa
This may be interpreted as the three-index version of paulions. Further, any operator on $\mathcal S$ commutes with any operator on $\mathcal B$.

Any state in $\mathcal H$ may be written as
\beq
\ket {\mathbf n_s} \otimes \ket {\mathbf n_b} \in \mathcal H,\quad \textnormal{with} \quad \ket{\mb 0}_{\mathcal H}\equiv \ket{\mb 0} = \ket{\mb 0}_{\mathcal S}\otimes \ket{\mb 0}_{\mathcal B}.
\eeq
The operator relations in (\ref{bowpaulions}) alone are not sufficient to encode the required exclusion statistics for arc diagrams (see equation (\ref{nconditions})). Indeed, any individual site on the arc diagram cannot be occupied by more than one arc foot. We therefore need to restrict ourselves to a particular subspace of $\mathcal H$. To this end we define the composite operators
\beq
c_{i,j,\sigma}=a_i a_j b_{i,j,\sigma}
\label{cij}
\eeq
and their adjoints
\beq
c_{i,j,\sigma}^\dagger = a_i^\dagger a_j^\dagger b_{i,j,\sigma}^\dagger.
\label{cijdagger}
\eeq
The action on the vacuum of $\mathcal H$ is defined as
\beq
c_{i,j,\sigma}^\dagger  \ket{\mb 0} = \ket{0,\ldots,n_i=1,0,\ldots,n_j=1,\ldots,0}\otimes\ket{0,\ldots, n_{i,j,\sigma}=1,0,\ldots,0}.
\eeq
Clearly the operators in equations (\ref{cij}) and (\ref{cijdagger}) are only non-zero if $i\neq j$, due to the paulionic nature of the $a$s. For the same reason it is further evident that for $k,j>i$, $c_{i,j,\sigma}^\dagger c_{i,k,\sigma}^\dagger = 0$, with a corresponding condition holding for the annihilation operators. Avoiding explicit ordering of indices, it holds in general that \textit{the product of any two $c^\dagger$s that share a position label is zero; similarly for the $c$s}. It is this property of the compound operators (\ref{cij}) and (\ref{cijdagger}) that captures the important restrictions on occupation numbers for arc diagrams set out in (\ref{nconditions}). We reiterate that the operator relations (\ref{bowpaulions}) alone would not be sufficient to achieve this. To see this we define super-occupation numbers,
\beq
N_{i,j,\sigma} = n_i \, n_j \, n_{i,j,\sigma},
\label{Nijdef}
\eeq
which count the compound species created by the operators in (\ref{cijdagger}): there must be an $a$-particle at \textit{each} site $i$ and $j$, \textit{and} there must be an arc of species $\sigma$ between $i$ and $j$ in order for $N_{i,j,\sigma} =1$. This set of super-occupation numbers obeys the conditions (\ref{nconditions}). Single-site occupation numbers $n_i$ label \textit{some} arc-foot at site $i$ on the arc diagram, where the other end of the arc could be at \textit{any other} site. The arc occupation numbers $n_{i,j,\sigma}$ label an arc of species $\sigma$ between sites $i$ and $j$. Thus if $N_{i,j,\sigma} =1$, there must be an arc-foot at each site $i$ and $j$, and there must be an arc (of species $\sigma$) connecting sites $i$ and $j$.

To the set $\Lambda$ of occupation numbers for a given arc diagram (as in equation (\ref{lambda})), we now associate the occupation number state
\beq
\left| \Lambda \right\rangle = \prod_{i,j,\sigma} (c_{i,j,\sigma}^\dagger)^{N_{i,j,\sigma}}\,\left| \mb 0 \right\rangle.
\label{lambdastate}
\eeq
This is in analogy with (\ref{nstate}). We further define a time-dependent state vector as in \eref{phit},
\beq
\ket{\phi(t)} \equiv \sum_{\Lambda} P(\Lambda|t) \ket{\Lambda},
\label{phitlambda}
\eeq
from which the probabilty of a particular configuration $\Lambda'$ may be obtained through
\beq
P(\Lambda'|t) = \overlap{\Lambda'}{\phi(t)}.
\label{probfromphilambda}
\eeq
\Eref{probfromphilambda} again provides the link between the master equation and the operator representation thereof, as set out in Section \ref{doisection1}. This enables us to write down the Liouvillian $\hat L$ for a given physical process so that formally
\beq
\ket{\phi(t)} = e^{\hat L t}\ket{\phi(0)}
\label{phitbow}
\eeq 
for some initial state $\ket{\phi(0)}$ of the system. Lastly we require a sum state on $\mathcal H$. In analogy to equation (\ref{projstate}), this is a uniform superposition of all possible occupation number states,
\beq
\ket s =  e^{\sum_i a_i^\dagger + \sum_{i,j,\sigma} b^\dagger_{i,j,\sigma}} \ket{\mb 0}.
\label{bowsum1}
\eeq
Given a Liouvillian, we can describe the time-evolution of quantities $\hat A$ as in equation (\ref{dttimeave}),
\beq
\partial_t \langle A(t) \rangle = \matrel{s}{[\hat A, \hat L]}{\phi(t)}.
\label{dttimeavebow}
\eeq
Here the quantity $A$ could be some function of the super-occupation numbers $N_{i,j,\sigma}$, the arc-occupation numbers $n_{i,j,\sigma}$ or the single-site occupation numbers $n_i$. Correspondingly the operator $\hat A$ would then be a function of the respective number operators
\beq
\hat N_{i,j,\sigma} \equiv c^\dagger_{i,j,\sigma}c_{i,j,\sigma},\quad \hat n_{i,j,\sigma}\equiv b^\dagger_{i,j,\sigma}b_{i,j,\sigma}, \quad \textnormal{and} \quad \hat n_i \equiv a^\dagger_i a_i.
\label{numberops}
\eeq
This equips us to describe dynamical behaviour of densities and correlators of crossings on arc-segments.

\subsection{The physical subspace \texorpdfstring{$\mathcal P \subset \mathcal H$}{P of H} and physical sum state}

The ``creation of arcs'' on the vacuum state of $\mathcal H$ with the composite operators $c^\dagger_{i,j,\sigma}$ according to(\ref{cijdagger}) and (\ref{lambdastate}) ensures that the requirements for valid arc diagrams set out in equation (\ref{nconditions}) are encoded into the physical states. We define the subspace $\mathcal P \subset \mathcal H$ of ``physical'' states as the span of all possible states created by the composite operators from the vacuum in $\mathcal H$. For instance, $a_i^\dagger b^\dagger_{i,j,\sigma} \ket{\mathbf 0}\notin \mathcal P$, since the $j$ index on the $b^\dagger$ is not paired with a corresponding $a_j^\dagger$.

If we assume that $\ket{\phi(0)}\in\mathcal P$, and that the Liouvillian leaves $\mathcal P$ invariant, then \eref{phitbow} implies $\ket{\phi(t)}\in \mathcal P$. It is thus sufficient to restrict our sum state (\ref{bowsum1}) to physical states,
\beq
\ket{s _{\mathcal P}} =  e^{\sum_{i,j,\sigma} c^\dagger_{i,j,\sigma}} \ket{\mb 0}.
\label{bowsum2}
\eeq
Instead of (\ref{dttimeavebow}) we may therefore write
\beq
\partial_t \langle A(t) \rangle = \matrel{s_{\mathcal P}}{[\hat A, \hat L]}{\phi(t)}.
\label{dttimeavebow2}
\eeq
%Note that the set theoretic complement of $\mathcal P$, denoted as $\bar{\mathcal P}$ is \textit{not} a vector space. To see this, take states $\phi_1 \in \mathcal P$ and $\phi_2 \in\bar{\mathcal P}$. It is clear that the states $\phi_\pm = \phi_1 \pm \phi_2$ lie in $\bar{\mathcal P}$, but that $\phi_+ + \phi_-$ lies in $\mathcal P$. Thus $\bar{\mathcal P}$ is not a vector space. Instead, consider the orthogonal complement of $\mathcal P$, denoted as $\mathcal P_\perp$, containing all states in $\mathcal H$ that are orthogonal to those in $\mathcal P$. $\mathcal P_\perp$ is indeed a vector space. This distinction is important when considering the form of the identity operator on $\mathcal H$ which may be written as the sum of projectors onto $\mathcal P$ and $\mathcal P_\perp$,
%\beq
%\hat{\mathbb I} _{\mathcal H} = \hat{\mathbb P}_{\mathcal P} + \hat{\mathbb P}_{\mathcal P _\perp}.
%\label{identityproj}
%\eeq
%The requirement that a given $\hat L$ should leave $\mathcal P$ invariant thus implies that $[\hat L,\hat{\mathbb P}_{\mathcal P}] = 0$. Since it is not trivial to write down the projector $\hat{\mathbb P}_{\mathcal P}$ explicitly, we shall simply test the analogous condition that all Liouvillians obey
Use of this equation for time evolution therefore requires that $\hat L$ must leave $\mathcal P$ invariant,
\beq
\hat L \ket{\phi} \in \mathcal P \quad \forall \ket \phi \in \mathcal P.
\label{invariant}
\eeq
Lastly, all Liouvillians must be probability conserving. Assuming that condition (\ref{invariant}) is met, and $\ket{ \phi(0)}\in \mathcal P$, probability conservation is encoded in terms of the \emph{physical} sum state (\ref{bowsum2}) through
\beq
\bra{s_{\mathcal P}} \hat L = 0.
\label{probconsbow}
\eeq

\subsection{Important relations of occupation numbers and properties of the physical subspace} %\texorpdfstring{$\mathcal P $}{P}}
\label{physicalprop}
We have defined the physical subspace $\mathcal P$ as the span of all states created by the action of the compound paulionic creation operators in \eref{cijdagger} on the vacuum in $\mathcal H$. States of this form --- see \eref{lambdastate} --- are labelled with super-occupation numbers $N_{i,j,\sigma}$ as defined in \eref{Nijdef}. We consider now some properties for relating super-occupation numbers, arc-occupation numbers and single site occupation numbers of any physical state, thereby encoding properties of arc diagrams.
\begin{enumerate} 
\item $N_{i,j,\sigma} = n_i \, n_j \,n_{i,j,\sigma}$ with $n_i, \, n_j, \,n_{i,j,\sigma} \in \{0,1\}$,
\item $n_{i,j,\sigma} = 1 \implies n_i = n_j = 1$ since if $\exists$ an arc between $i$ and $j$ there must be arc feet at these sites,
\item $n_{i,j,\sigma} = 0 \notimplies n_i = 0 \;\;\textnormal{or} \;\; n_j = 0$ since the absence of an arc between $i$ and $j$ does not imply that feet of other arcs may not be at these sites,
\item $n_i = 0 \;\;\textnormal{and / or} \;\; n_j = 0 \implies n_{i,j,\sigma}=0$ since the absence of arc feet at sites $i$ and / or $j$ implies that there also cannot be an arc between them, and
\item $n_i = 1\;\;\textnormal{and / or} \;\; n_j = 1 \notimplies n_{i,j,\sigma}=1$ since the presence of arc feet at sites $i$ and $j$ does not imply that there is an arc between them.
\end{enumerate}
Consequently it is clear that $N_{i,j,\sigma} = n_{i,j,\sigma}$. By considering the possible different combinations, one may also conclude that, for instance,
\beq
n_i \, n_{i,j,\sigma} = n_{i,j,\sigma}.
\label{ninij}
\eeq
This equivalence also holds on the level of the corresponding number operators (\ref{numberops}). Therefore some operators are equivalent on $\mathcal P$. For instance, instead of checking whether there exists some arc that has a foot at site $i$, one could simply check whether site $i$ is occupied by \textit{some} arc foot, i.e., $a^\dagger_i a_i = \sum_{\sigma} \left\{ \sum_{k(<i)}  b^\dagger_{k,i,\sigma}\,b_{i,k,\sigma} + \sum_{k(>i)} b^\dagger_{i,k,\sigma}\,b_{i,k,\sigma} \right\} = \sum_{\sigma} \left\{ \sum_{k(<i)}  c^\dagger_{k,i,\sigma}\,c_{i,k,\sigma} + \sum_{k(>i)} c^\dagger_{i,k,\sigma}\,c_{i,k,\sigma} \right\}$. Several other examples exist, but all are based on the above properties. It is further useful to note the following properties of expectation values of the form $\langle \cdot \rangle = \matrel{s_{\mathcal P}}{\cdot}{\phi(t)}$:
\begin{enumerate}
\item $\langle c_{i,k,\sigma} \rangle = \langle  c^\dagger_{i,k,\sigma} c_{i,k,\sigma} \rangle = \langle  \hat N _{i,k,\sigma} \rangle$ since $\bra{s_{\mathcal P}}c_{i,k,\sigma}=\bra{s_{\mathcal P}}c^\dagger_{i,k,\sigma}c_{i,k,\sigma}$ (in analogy to \eref{nons}),
\item similarly $\langle c^\dagger_{i,k,\sigma} \rangle = \langle  c_{i,k,\sigma}  c^\dagger_{i,k,\sigma} \rangle $,
\item $\langle a_i \rangle \neq \langle a^\dagger_i a_i \rangle$ since $\bra{s_{\mathcal P}} a_i \neq \bra{s_{\mathcal P}} a_i^\dagger a_i$,
\item $\langle  \hat N _{i,k,\sigma} c_{i,j,\sigma'} \rangle=\langle   a_i^\dagger a_i a_k^\dagger a_k b^\dagger_{i,k,\sigma} b_{i,k,\sigma} a_i a_j b_{i,j,\sigma'} \rangle=0$ $\forall j,k,\sigma,\sigma'$ since $a_i^2=0$,
\item similarly $\langle c_{i,j,\sigma'} \hat N _{i,k,\sigma}  \rangle = \langle  \hat N _{i,k,\sigma} c^\dagger_{i,j,\sigma'} \rangle=\langle   c^\dagger_{i,j,\sigma'} \hat N _{i,k,\sigma}  \rangle=0$, due to the states that the number operators select and the $c$s and $c^\dagger$s annihilate or create when acting on $\bra{s_{\mathcal P}}$, 
\item $\langle c_{i,j,\sigma} c^\dagger_{i,j,\sigma} \rangle = \langle a_i a_i^\dagger a_j a_j^\dagger  b_{i,j,\sigma}b^\dagger_{i,j,\sigma} \rangle = \langle (1-\hat n_i)(1 - \hat n_j)(1- \hat n_{i,j,\sigma}) \rangle \neq \langle 1-\hat N_{i,j,\sigma}\rangle$, and
\item $\langle (1-\hat n_i)(1 - \hat n_j)(1- \hat n_{i,j,\sigma}) \rangle = \ave{(1-\hat n_i)(1 - \hat n_j)} $ due to equation \eref{ninij}.
\end{enumerate}
These properties will be needed for calculations involving equation \eref{dttimeavebow2}.

\subsection{\textbf{R0}: Liouvillian and dynamical quantities}
\label{R0doisect}
Here we shall describe the move \textbf{R0} on arc diagrams (see Section \ref{R0descript}) in terms of the operator formalism set out above. We construct the Liouvillian corresponding to this process, drawing on the example of single-species paulionic diffusion set out in equation (\ref{Ldiffpaul}) where diffusion of a particle from site $i$ to site $j$ was considered. In the context of arc diagrams, we have in mind an arc whose one foot is at site $i$, and then ``diffuses'' to site $j$. The other foot of the arc, at site $k$, is not involved in this process and remains stationary. The corresponding Liouvillian is
\beqa
\hat L _{\textnormal R0} &=& D\sum_{<i,j>} \sum_\sigma \bigg\{ \sum_{k=1}^{\textrm{min}(i,j)} \;\;\left[ c^\dagger_{k,j,\sigma}c_{k,i,\sigma} - a_ja^\dagger_j \, c^\dagger_{k,i,\sigma}c_{k,i,\sigma} \right] \nl
&& \qquad \qquad\;+ \sum_{k=\textrm{max}(i,j)}^{N} \left[ c^\dagger_{j,k,\sigma}c_{i,k,\sigma} - a_ja^\dagger_j \, c^\dagger_{i,k,\sigma}c_{i,k,\sigma} \right] \bigg\},
\label{LR0}
\eeqa
where $D$ is some diffusion constant. The outer summation runs over nearest neighbours $i$ and $j$, and the two summations over $k$ account for the two cases $k<i,j$ and $k>i,j$. The latter summations encode that the \textit{same} arc undergoes diffusion of its one foot, and includes all possible arcs that could undergo this step. In both cases the operator $a_ja^\dagger_j  = 1 - a^\dagger_j a_j$ allows diffusion to occur only if the target site is unoccupied, as with equation (\ref{Ldiffpaul}). It is indeed sufficient to perform this check on the single-site occupancy number $n_j$, since $n_j = 0 \implies N_{k,j,\sigma}=0$ $\forall k$ (see \eref{nconditions} and \sref{physicalprop}). The interpretation of the other terms is clear: the positive term ``picks out'' a pre-cursor state where an arc, whose one foot is based at site $k$, has its other arc foot at site $i$, but not at site $j$. The negative term indicates that the diffusion step out of the current state may only occur if $j$ is unoccupied and there is some arc with a foot at site $i$. The Liouvillian in equation \eref{LR0} leaves the physical subspace $\mathcal P$ invariant since all contributions that add or remove particles (arc feet) are written in terms of $c$s and $c^\dagger$s. It remains to check that $\hat L _{\textnormal{R0}}$ is probability conserving, i.e., that $\bra{s_{\mathcal P}}\hat L _{\textnormal{R0}} = 0$ or equivalently $\hat L _{\textnormal{R0}}^\dagger \ket{s_{\mathcal P}} = 0$. Consider the action of the adjoint of the operators in the first line of \eref{LR0} on the physical sum state.
%\beq
%\big[ \underbrace{c_{k,i,\sigma}^\dagger c_{k,j,\sigma}}_{\equiv \hat A} - \underbrace{a_ja^\dagger_j \, c^\dagger_{k,i,\sigma}c_{k,i,\sigma}}_{\equiv \hat B} \big] \ket{s_{\mathcal P}}.
%\label{LR0probcons}
%\eeq
For instance, $c_{k,i,\sigma}^\dagger c_{k,j,\sigma}$ annihilates all terms in $\ket{s_{\mathcal P}}$ except the ones for which $N_{k,j,\sigma}=1$, setting $N_{k,j,\sigma}=0$ in these terms. Of the remaining terms, all are annihilated except those where $N_{k,i,\sigma}=0$, and then $N_{k,i,\sigma}=1$ is enforced. What remains are all terms where $N_{k,j,\sigma}=0$ and $N_{k,i,\sigma}=1$. The operator $a_ja^\dagger_j \, c^\dagger_{k,i,\sigma}c_{k,i,\sigma}$ annihilates all terms in $\ket{s_{\mathcal P}}$ where $n_j=1$ and also all terms where $N_{k,i,\sigma}=0$, leaving those where $n_j=0$ and $N_{k,i,\sigma}=1$. However, if $N_{k,i,\sigma}=1$, the two statements $n_j=0$ and $N_{k,j,\sigma}=0$ are equivalent for any state in $\mathcal P$ since $N_{k,j,\sigma}=0\implies n_j =0$, and conversely $(N_{k,i,\sigma}=1$ and $n_j=0)\implies N_{k,j,\sigma}=0 $; see again \eref{nconditions}. Similar reasoning applies to the other terms in \eref{LR0}, and thus this $\hat L _{\textnormal{R0}} $ is probability conserving, as required.

We now calculate some dynamical quantities using the machinery set out in Section \ref{doisectionmain}, and therefore repeat here the time-evolution equation for dynamical quantities \eref{dttimeavebow2},
\beq
\partial_t \langle A(t) \rangle = \matrel{s_{\mathcal P}}{[\hat A, \hat L]}{\phi(t)} = \langle [\hat A, \hat L] \rangle.
\label{r0timeevol}
\eeq

We begin with the single-site number operator $\hat n_I = a_I^\dagger a_I$ for a particular site $I$. This operator clearly commutes with the negative terms of the Liouvillian for \textbf{R0} in \eref{LR0}. For $k<i,j$, for instance, we have 
\beq
[\hat n_I, a_ja^\dagger_j \, c^\dagger_{k,i,\sigma}c_{k,i,\sigma} ] = 0.
\eeq
The non-trivial commutators are with the positive terms in \eref{LR0}, e.g.,
\beqa
[\hat n_I, c^\dagger_{k,j,\sigma}c_{k,i,\sigma}  ] &=& a_k^\dagger a_k b^\dagger_{k,j,\sigma}b_{k,i,\sigma} \;[a^\dagger_I a_I,a_j^\dagger a_i]\nl
&=& a_k^\dagger a_k b^\dagger_{k,j,\sigma}b_{k,i,\sigma} \;\big\{ a_j^\dagger [a^\dagger_I a_I,a_i] + [a^\dagger_I a_I,a_j^\dagger] a_i \big\}.
\label{nIR0a}
\eeqa
For the case $k>i,j$ the position labels on the $b$ and $b^\dagger$ would be exchanged. Using the paulionic commutation relations in equation \eref{paulions} we find that
\beqa
[a^\dagger_\alpha a_\alpha,a_\beta]&=& \delta_{\alpha,\beta}\,(2a^\dagger_\alpha a_\alpha-1)\,a_\alpha = -\delta_{\alpha,\beta}\,a_\alpha,\quad \textnormal{and} \nl
\,[a^\dagger_\alpha a_\alpha,a_\beta^\dagger]&=& a_\alpha^\dagger\,\delta_{\alpha,\beta}\,(1-2a^\dagger_\alpha a_\alpha)=\delta_{\alpha,\beta}\,a_\alpha^\dagger.
\eeqa
Inserting this into \eref{nIR0a} and then performing cancellations and relabellings in the summations of the Liouvillian (\ref{LR0}) in the time-evolution equation yields
\beqa
\partial_t \langle \hat n_I \rangle_{\textnormal R0} &=& D\sum_\sigma \sum_{i(I)} \bigg\{ \sum_{k=1}^{\textrm{min}(i,I)}\;\, \left(\langle c^\dagger_{k,I,\sigma}c_{k,i,\sigma}\rangle - \langle c^\dagger_{k,i,\sigma}c_{k,I,\sigma}\rangle\right) \nl
&&\qquad\quad\;\; + \sum_{k=\textrm{max}(i,I)}^{N} \left( \langle c^\dagger_{I,k,\sigma}c_{i,k,\sigma}\rangle - \langle c^\dagger_{i,k,\sigma}c_{I,k,\sigma}\rangle \right) \bigg\}.
\label{nIR0b}
\eeqa
Here $i(I)$ are again sites $i$ that are nearest neighbours of $I$. Let us consider a generic term in this expression,
\beqa
\langle c^\dagger_{k,I,\sigma}c_{k,i,\sigma}\rangle &=& \langle (1-\hat n_k)(1 - \hat n_I)(1- \hat n_{k,I,\sigma}) c_{k,i,\sigma}\rangle \nl
&=& \underbrace{\ave{c_{k,i,\sigma}}}_{\ave{\hat N_{k,i,\sigma}}} -\underbrace{\ave{\hat n_k c_{k,i,\sigma}}}_0- \ave{\hat n_I c_{k,i,\sigma}} -\ave{\hat n_{k,I,\sigma} c_{k,i,\sigma}} \nl
&&+\underbrace{\ave{\hat n_k \hat n_I c_{k,i,\sigma}}}_0 +\underbrace{\ave{\hat n_k \hat n_{k,I,\sigma} c_{k,i,\sigma}}}_0 + \ave{c_{ k,i,\sigma}\underbrace{\hat n_I \hat n_{k,I,\sigma}}_{\hat n_{k,I,\sigma}}}- \underbrace{\ave{\hat N_{k,I,\sigma}c_{k,i,\sigma}}}_0 \nl
&=& \ave{\hat N_{k,i,\sigma}(1- \hat n_I)} \nl
&=& \ave{\hat n_{k,i,\sigma}(1- \hat n_I)}
\label{cdaggercave}
\eeqa
where we have used the fact that $i\neq I$ and also the properties set out in Section \ref{physicalprop}. We can now do the summations over $k$ and $\sigma$ explicitly in \eref{nIR0b}, since we know that $\sum_\sigma (\sum_{k(<i)}\hat n_{k,i,\sigma}+\sum_{k(>i)}\hat n_{i,k,\sigma})=\hat n_i$. We obtain the simple relation 
\beq
\partial_t \ave{\hat n_I}_{\textnormal R0} = D \sum_{i(I)} \left[ \ave{\hat n_i} - \ave{\hat n_I} \right],
\label{nIR0c}
\eeq
which is exactly the same discrete diffusion equation obtained in equation \eref{dtnIbos} for diffusion of a single species of paulions or bosons. (The right side above is simply the discrete version of the second order spatial derivative of the density.) We conclude the following important result: the diffusion of arc-feet is a local process, which is insensitive to the location of the non-diffusing arc-feet.

We now calculate the single-site correlator,
\beq
\partial_t \langle \hat n_I \hat n_J \rangle_{\textnormal R0} =  \langle [\hat n_I \hat n_J, \hat L_{\textnormal{R0}}] \rangle.
\eeq
Consequently we require commutators of the following type,
\beqa
[\hat n_I \hat n_J, c^\dagger_{k,j,\sigma}c_{k,i,\sigma}  ] &=& a_k^\dagger a_k b^\dagger_{k,j,\sigma}b_{k,i,\sigma} \;[a^\dagger_I a_I\,a^\dagger_J a_J,a_j^\dagger a_i]\nl
&=& a_k^\dagger a_k b^\dagger_{k,j,\sigma}b_{k,i,\sigma} \;\big\{ -\delta_{i,J} a_j^\dagger \hat n_I a_i - \delta_{i,I} a_j^\dagger  a_i \hat n_J + \delta_{j,J} \hat n_I a_j^\dagger a_i + \delta_{j,I} a_j^\dagger \hat n_J a_i \big\} \nl
&=& c^\dagger_{k,j,\sigma}c_{k,i,\sigma} \bigg\{ \delta_{i,J}\,\delta_{i,I} + \delta_{j,J}\,\delta_{j,I} -\delta_{j,J}\,\delta_{i,I} - \delta_{j,I}\,\delta_{i,J} \nl
&&\quad + \hat n_I \left(\delta_{j,J}-\delta_{i,J}  \right) + \hat n_J \left(\delta_{j,I}-\delta_{i,I}  \right) \bigg\}.
\eeqa
We note that, since $I\neq J$, terms such as $\delta_{i,J}\,\delta_{i,I}$ vanish. Using equation \eref{cdaggercave} we conclude that 
\beqa
\partial_t \langle \hat n_I \hat n_J \rangle_{\textnormal R0} &=& D\sum_{k\neq i,j} \sum_\sigma \bigg\{ -\delta_{<I,J>} \left( \ave{\hat n_{k,I,\sigma}(1-\hat n_J )}+\ave{\hat n_{k,J,\sigma}(1-\hat n_I )} \right) \nl
&& \quad + \sum_{i(J)}\ave{\hat n_{k,i,\sigma}(1-\hat n_J )\hat n_I} + \sum_{j(J)}\ave{\hat n_{k,J,\sigma}(1-\hat n_j )\hat n_I}\nl
&& \quad + \sum_{i(I)}\ave{\hat n_{k,i,\sigma}(1-\hat n_I )\hat n_J} + \sum_{j(I)}\ave{\hat n_{k,I,\sigma}(1-\hat n_j )\hat n_J} \bigg\}.
\eeqa
In the summation over $k$ we have implied the correct ordering of indices on the $\hat n$s according to $k<i,j$ or $k>i,j$. Using the properties from Section \ref{physicalprop} this summation and that over $\sigma$ may be done explicitly, yielding
\beqa
\partial_t \langle \hat n_I \hat n_J \rangle_{\textnormal R0} &=& D \; \bigg\{ -\delta_{<I,J>} \left( \ave{\hat n_I}  + \ave{\hat{n_J}} \right) \nl
&& \quad +\sum_{i(J)} \left( \ave{\hat n_i \hat n_I} - \ave{\hat n_J \hat n_I} \right) \nl
&& \quad +\sum_{i(I)} \left( \ave{\hat n_i \hat n_J} - \ave{\hat n_I \hat n_J} \right) \bigg\}.
\eeqa
For the case that $I$ and $J$ are not nearest neighbours, a continuum version of this equation would could be written as $\partial_t \, c(x,y) = D' (\frac{\partial^2}{\partial x^2} c + \frac{\partial^2}{\partial y^2} c)$ where $D'$ is some rescaled diffusion constant arising from the continuum limit. If $I$ and $J$ are nearest neighbours, one simply obtains discrete versions of the gradient.

Next we turn to the number operator for arcs, $\hat n_{I,J,\tilde\sigma}= b^\dagger_{I,J,\tilde\sigma} b_{I,J\tilde\sigma}$. This operator also commutes with the negative terms in Liouvillian \eref{LR0}. In analogy to equation \eref{nIR0a}, non-trivial commutators of the following type remain,
\beqa
[\hat n_{I,J,\tilde\sigma}, c^\dagger_{k,j,\sigma}c_{k,i,\sigma}  ] &=& a_k^\dagger a_k a_j^\dagger a_i  \;[b^\dagger_{I,J,\tilde\sigma} b_{I,J\tilde\sigma},b^\dagger_{k,j,\sigma}b_{k,i,\sigma}] \nl
&=& a_k^\dagger a_k a_j^\dagger a_i  \;\left\{ b^\dagger_{k,j,\sigma}[\hat n_{I,J,\tilde\sigma},b_{k,i,\sigma}] + [\hat n_{I,J,\tilde\sigma},b^\dagger_{k,j,\sigma}] b_{k,i,\sigma}\right\}.
\eeqa
We may now use the relations \eref{bowpaulions} to derive the following condition,
\beqa
[\hat n_{\alpha,\beta,\sigma},b_{\gamma,\delta,\sigma'}]&=&  -\delta_{\alpha,\gamma}\,\delta_{\beta,\delta}\,\delta_{\sigma,\sigma'}\,b_{\alpha,\beta,\sigma},\quad \textnormal{and} \nl
\,[\hat n_{\alpha,\beta,\sigma},b^\dagger_{\gamma,\delta,\sigma'}]&=& \delta_{\alpha,\gamma}\,\delta_{\beta,\delta}\,\delta_{\sigma,\sigma'}\,b^\dagger_{\alpha,\beta,\sigma}.
\eeqa
This may be re-inserted into the time-evolution equation for $\hat n_{I,J,\tilde\sigma}$, and simplified through equation \eref{cdaggercave} to obtain
\beqa
\partial_t \langle \hat n_{I,J,\tilde\sigma} \rangle_{\textnormal R0} &=& D\sum_\sigma \sum_{<i,j>} \bigg\{ \sum_{k=1}^{\textrm{min}(i,j)}\;\, \left( \delta_{I,k}\;\delta_{\sigma,\tilde\sigma} \right) \left( \delta_{K,j} - \delta_{K,i} \right)\ave{\hat N_{k,i,\sigma}(1-\hat n_j)} \nl
&&\qquad\quad\;\;\;\; + \sum_{k=\textrm{max}(i,I)}^{N} \left( \delta_{I,k}\;\delta_{\sigma,\tilde\sigma} \right) \left( \delta_{K,j} - \delta_{K,i} \right)\ave{\hat N_{i,k,\sigma}(1-\hat n_j)} \bigg\}.
\eeqa
The first line of this equation (the case where $k<i,j$) could be simplified as follows,
\beqa
&&\sum_\sigma \sum_{<i,j>} \sum_{k(<i,j)}\delta_{I,k}\,\delta_{\sigma,\tilde\sigma} \left( \delta_{J,j} - \delta_{J,i} \right)\ave{\hat N_{k,i,\sigma}(1-\hat n_j)}\nl
&&= \sum_{<i,j>} \left( \delta_{J,j} - \delta_{J,i} \right)\ave{\hat N_{I,i,\tilde\sigma}(1-\hat n_j)}\nl
&&= \sum_{i(J)}\ave{\hat N_{I,i,\tilde\sigma}(1-\hat n_J)} - \sum_{j(J)}\ave{\hat N_{I,J,\tilde\sigma}(1-\hat n_j)} \nl
&& = \sum_{i(J)} \left( \ave{\hat N_{I,i,\tilde\sigma}(1-\hat n_J)} - \ave{\hat N_{I,J,\tilde\sigma}(1-\hat n_i)} \right).
\label{r0bowcora}
\eeqa
Again recalling that $N_{i,j,\sigma}=n_{i,j,\sigma}$ and multiplying out the terms above, we conclude that for the case $k<i,j$
\beq
\partial_t \langle \hat n_{I,J,\tilde\sigma} \rangle_{\textnormal R0} = D\sum_{i(J)} \bigg( \ave{\hat n_{I,i,\tilde\sigma}} - \ave{\hat n_{I,J,\tilde\sigma}} \bigg) + 
D\sum_{i(J)} \bigg( \ave{\hat n_{I,J,\tilde\sigma}\hat n_i} - \ave{\hat n_{I,i,\tilde\sigma} \hat n_J} \bigg).
\eeq
The first summation is simply a discrete diffusion equation for the right foot of an arc where the left one is kept fixed --- compare to the single-site diffusion equation \eref{nIR0c}. The second summation shows that arc diffusion also involves a correlation between the diffusing arcs and the single-site occupancies. The result for $k>i,j$ is trivially obtainable in a similar manner. 

Lastly we calculate the correlator for arc occupancies.  We shall require the following commutator,
\beqa
[\hat n_{I,J,\sigma_1} \,\hat n_{K,L,\sigma_2}\,,\, c^\dagger_{k,j,\sigma}c_{k,i,\sigma}  ] &=& a_k^\dagger a_k a_j^\dagger a_i  \;[\hat n_{I,J,\sigma_1} \,\hat n_{K,L,\sigma_2}\,,\,b^\dagger_{k,j,\sigma}b_{k,i,\sigma}] \nl
&=& c^\dagger_{k,j,\sigma}c_{k,i,\sigma} \;\Big\{ \big( \delta_{k,K}\,\delta_{j,L}\,\delta_{\sigma,\sigma_2} -\delta_{k,K}\,\delta_{i,L}\,\delta_{\sigma,\sigma_2}\big)\hat n_{I,J,\sigma_1} \nl
&&\;\;\;+ \;\;\;\big( \delta_{k,I}\,\delta_{j,J}\,\delta_{\sigma,\sigma_1} - \delta_{k,I}\,\delta_{i,J}\,\delta_{\sigma,\sigma_1}\big)\hat n_{K,L,\sigma_2} \Big\}.
\eeqa
Inserting this into the time-evolution equation \eref{r0timeevol} with the Liouvillian \eref{LR0} we obtain
\beqa
\frac {\partial_t \langle \hat n_{I,J,\sigma_1} \hat n_{K,L,\sigma_2} \rangle_{\textnormal R0}} D  &=& \sum_{i(L)} \Big[ \ave{\hat n_{K,i,\sigma_2}(1-\hat n_L)\hat n_{I,J,\sigma_1}} - \ave{\hat n_{K,L,\sigma_2}(1-\hat n_i)\hat n_{I,J,\sigma_1}} \Big] +\;\;\nl
&&  \sum_{i(J)} \Big[ \ave{\hat n_{I,i,\sigma_1}(1-\hat n_J)\hat n_{K,L,\sigma_2}} - \ave{\hat n_{I,J,\sigma_1}(1-\hat n_i)\hat n_{K,L,\sigma_2}} \Big].
\eeqa
Comparing this with the arc-diffusion equation (\ref{r0bowcora}), we note that $\hat n_{I,J,\sigma_1}$ is only correlated with $\hat n_{K,L,\sigma_2}$ if the arc between sites $K$ and $L$ is undergoing diffusion (and vice versa). Since we are considering only the \textbf{R0}-move in this correlator, the expected encoding of the diffusive behaviour is manifest here. This result would lend itself well to a mean field approximation for decoupling arc diffusion terms of the type \eref{r0bowcora} and arc densities.

\subsection{Boundary conditions on arc diagrams}

The periodic boundary condition set out in Section \ref{bowdiagsect} (see Figure \ref{fig:bowboundcond}) must be encoded into the \textbf{R0} process. Diffusion of an arc foot across this boundary must result in a sign change of the arc. This could be done by augmenting the Liouvillian for \textbf{R0} with a sign-changing term at the boundary. Alternatively, quantities can be calculated ``in the bulk'' (i.e., away from this boundary), and a corresponding boundary current term can be included by hand.

\subsection{\textbf{R1}: Liouvillian and dynamical quantities}
\label{R1doisect}
Reidemeister 1 involves the creation and annihilation of a single arc at nearest-neighbour sites on the line, as stated in Section \ref{R1descript} (see Figure \ref{fig:r1bow}). The Liouvillian for the arc-creation process is 
\beq
\hat L _{\textnormal{R1,cr.}} = g\sum_{i=1}^N\sum_{\sigma} \bigg\{c^\dagger_{i,i+1,\sigma} - c_{i,i+1,\sigma}c^\dagger_{i,i+1,\sigma}\bigg\},
\label{LR1cr}
\eeq
where $g$ is some rate constant. The analogies to the paulionic Liouvillian for single-species particle creation \eref{Lcrpaul} are clear. The positive term selects all precursor states which do not have an arc at nearest-neighbouring sites $i$ and $i+1$. The negative term ensures that flux out of the current state through creation of an arc may only happen if the current state has no arcs at these nearest neighbouring sites. $\hat L_{\textnormal{R1,cr.}}$ leaves $\mathcal P$ invariant, and can beshown to be propability conserving by considering the action of the adjoint of each operator in \eref{LR1cr} on the physical sum state.

The Liouvillian for the arc-annihilation process is 
\beq
\hat L _{\textnormal{R1,an.}} = h\sum_{i=1}^N\sum_{\sigma} \bigg\{c_{i,i+1,\sigma} - c^\dagger_{i,i+1,\sigma}c_{i,i+1,\sigma}\bigg\},
\label{LR1an}
\eeq
where $h$ is some rate constant. Here, too, the analogies to the corresponding paulionic Liouvillian for single-species particle annihilation \eref{Lanpaul} are evident. The positive term selects a precursor state with one more arc between sites $i$ and $i+1$ than the current state. The negative term enforces that the current state may only be exited through annihilation of an arc between these sites if indeed such an arc exists. It is easy to verify that the Liouvillian for arc-annihilation \eref{LR1an} also leaves the physical subspace $\mathcal P$ invariant and is probability conserving, as required.

Next we calculate dynamical quantities for the \textbf{R1} Liouvillians \eref{LR1cr} and\eref{LR1an}. As for \textbf{R0}, all number operators commute with the negative parts of both of these Liouvillians. Again we require the several commutators, obtained from the paulionic relations \eref{paulions} and \eref{bowpaulions}. For the single-site quantities we need the following,
\beqa
[\hat n_I, c^\dagger_{i,i+1,\sigma} ] &=& (\delta_{I,i}+\delta_{I,i+1})\; c^\dagger_{i,i+1,\sigma},\nl
\,[\hat n_I, c_{i,i+1,\sigma}  ] &=& -(\delta_{I,i}+\delta_{I,i+1}) \; c_{i,i+1,\sigma},\nl
\,[\hat n_I \hat n_J, c^\dagger_{i,i+1,\sigma}  ] &=&  c^\dagger_{i,i+1,\sigma}\; \Big\{\delta_{i+1,J} \left( \hat n_I + \delta_{i+1,I} + \delta_{i,I}\right) + \delta_{i,J} \left( \hat n_I + \delta_{i+1,I} + \delta_{i,I}\right)\nl
&& \qquad +\delta_{i+1,I}\,\hat n_J +\delta_{i,I}\,\hat n_J \Big\},\nl
\,[\hat n_I \hat n_J, c_{i,i+1,\sigma}  ] &=& c_{i,i+1,\sigma} \; \Big\{-\delta_{i+1,J} \left( \hat n_I + \delta_{i+1,I} - \delta_{i,I}\right) -\delta_{i,J} \left( \hat n_I - \delta_{i+1,I} - \delta_{i,I}\right)\nl
&& \qquad -\delta_{i+1,I}\,\hat n_J -\delta_{i,I}\,\hat n_J \Big\}.
\label{r1comms1}
\eeqa
For the arc quantities we will need the following additional commutators,
\beqa
\,[\hat n_{I,J,\tilde\sigma}\,,\, c^\dagger_{i,i+1,\sigma}  ] &=& \delta_{I,i}\,\delta_{J,i+1}\, \delta_{\sigma,\tilde\sigma}\;c^\dagger_{i,i+1,\sigma} ,\nl
\,[\hat n_{I,J,\tilde\sigma}\,,\, c_{i,i+1,\sigma} ] &=& -\delta_{I,i}\,\delta_{J,i+1} \,\delta_{\sigma,\tilde\sigma}\;c_{i,i+1,\sigma},\nl
\,[\hat n_{I,J,\sigma_1}\,\hat n_{K,L,\sigma_2}\,,\, c^\dagger_{i,i+1,\sigma}  ] &=& c^\dagger_{i,i+1,\sigma}\;\Big\{\delta_{I,i}\,\delta_{J,i+1}\, \delta_{\sigma_1,\sigma}\,\hat n_{K,L,\sigma_2} \nl
&& \quad + \delta_{K,i}\,\delta_{L,i+1}\, \delta_{\sigma_2,\sigma} \,\hat n_{I,J,\sigma_1}\Big\} ,\nl
\,[\hat n_{I,J,\sigma_1}\,\hat n_{K,L,\sigma_2}\,,\, c_{i,i+1,\sigma} ] &=& c_{i,i+1,\sigma}\;\Big\{-\delta_{I,i}\,\delta_{J,i+1}\, \delta_{\sigma_1,\sigma}\,\hat n_{K,L,\sigma_2} \nl
&& \quad - \delta_{K,i}\,\delta_{L,i+1}\, \delta_{\sigma_2,\sigma} \,\hat n_{I,J,\sigma_1}\Big\}.
\label{r1comms2}
\eeqa
Furthermore, $\ave{c_{i,i+1,\sigma} } = \ave{\hat N_{i,i+1,\sigma} }= \ave{\hat n_{i,i+1,\sigma} }$ and $\ave{c^\dagger_{i,i+1,\sigma} } = \ave{(1-\hat n_i)(1-\hat n_{i+1})}$. This, together with the commutators \eref{r1comms1} and \eref{r1comms2}, allows for calculating the time-evolution of several average densities. 

We begin with the single-site density. For the \textbf{creation of an arc} we obtain
\beq
\partial_t \ave{\hat n_{I}}_{\textnormal{R1,cr.}} = g \sum_\sigma \big[ \ave{(1-\hat n_I)(1-\hat n_{I+1})}+\ave{(1-\hat n_{I-1})(1-\hat n_{I})} \big]. 
\label{dtnir1cr}
\eeq
The interpretation here is clear: the creation of an arc can only increase the single-site density $\hat n_I$, and this can only happen if both the site $I$ and one of its neighbours are unoccupied. Saturation effects of the restricted occupancy are evident. Note that this differential equation only depends on single-site densities, and entails no coupling to arc densities.

The single-site correlator for the \textbf{R1} creation process is found to be
\beqa
\partial_t \ave{\hat n_{I}\hat n_{J}}_{\textnormal{R1,cr.}}&=& g\;\sum_\sigma \bigg\{ \delta_{<I,J>} \big[ \ave{(1-\hat n_{I})(1-\hat n_{J})} + \ave{(1-\hat n_{I})(1-\hat n_{I})}\big] \nl
&&\quad\;\; + \big[ \ave{(1-\hat n_{J-1})(1-\hat n_{J})\hat n_I} + \ave{(1-\hat n_{J})(1-\hat n_{J+1})\hat n_I} \big] \nl
&&\quad\;\;  + \big[ \ave{(1-\hat n_{I-1})(1-\hat n_{I})\hat n_J} + \ave{(1-\hat n_{I})(1-\hat n_{I+1})\hat n_J} \big] \bigg\}.
\label{ninjr1cr}
\eeqa
This correlator is a complicated function of higher orders of single-site correlators. The first term on the right indicates correlation if $I$ and $J$ are nearest neighbours. The remaining terms show that site occupancy at site $I$ is correlated with an unoccupied site $J$ and one of its unoccupied nearest neighbours, and vice-versa. This makes sense, since the process under consideration here is the creation of arcs, which happens at empty nearest-neighbouring sites. Again the saturation brought about by occupancy restrictions is evident. In equation (\ref{ninjr1cr}) one could, in principle, multiply out the various terms. A continuum version would then include gradient terms and triplet correlator terms.

We now repeat this analysis for the arc-density, obtaining
\beqa
\partial_t \ave{\hat n_{I,J,\tilde\sigma}}_{\textnormal{R1,cr.}} &=& g\sum_i \sum_\sigma \;\delta_{I,i}\,\delta_{J,i+1} \,\delta_{\sigma,\tilde\sigma} \;\ave{c^\dagger_{i,i+1,\sigma}}\nl
&=& g \;\delta_{<I,J>}\ave{(1-\hat n_I)(1-\hat n_{J})},
\eeqa
where the $\delta_{<I,J>}$ ensures that the arc-density $\hat n_{I,J,\tilde\sigma}$ can only increase through the creation of an arc if indeed $I$ and $J$ are nearest neighbours. It is further clear from the term $\ave{(1-\hat n_I)(1-\hat n_{J})}$ that the sites must be unoccupied in order for an arc to be created --- the restriction of occupation numbers is manifest. 

One may also calculate the correlator for arc quantities, 
\beqa
\partial_t \ave{\hat n_{I,J,\sigma_1}\,\hat n_{K,L,\sigma_2}}_{\textnormal{R1,cr.}} &=& g\; \Big\{\delta_{<I,J>}\ave{(1-\hat n_{I,J,\sigma_1})\hat n_{K,L,\sigma_2}} \nl
&& + \;\;\delta_{<K,L>}\ave{(1-\hat n_{K,L,\sigma_2})\hat n_{I,J,\sigma_1}}\Big\}.
\eeqa
This result may be interpreted as follows: under the \textbf{R1}-creation move, two arcs are only correlated if the occupancy conditions for a creation move are met. This is ensured by the ``$1-n$'' terms and the Kronecker delta functions.

Turning to the Liouvillian for the \textbf{annihilation process}, we obtain for the single-site density
\beq
\partial_t \ave{\hat n_{I}}_{\textnormal{R1,an.}} = -h \sum_\sigma \big[ \ave{\hat n_{I,I+1,\sigma}}+\ave{\hat n_{I-1,I,\sigma}} \big].
\eeq
The implication is that the single-site density $\hat n_I$ can only be decreased by the annihilation of an arc if there exists an arc (of either species) that has one foot at site $I$ and another foot at a nearest neighbouring site. In contrast to equation \eref{dtnir1cr}, this differential equation involves explicit dependence of a single-site quantity on the arc densities.

The single-site correlator for the \textbf{R1} annihilation process is found to be
\beqa
\partial_t \ave{\hat n_{I}\hat n_{J}}_{\textnormal{R1,an.}}&=& h\;\sum_\sigma \bigg\{ \delta_{<I,J>} \big[ \ave{\hat n_{I,J,\sigma}} + \ave{\hat n_{J,I,\sigma}} \big] \nl
&&\quad\quad - \big[ \ave{\hat n_{J-1,J,\sigma}\hat n_I} + \ave{\hat n_{J,J+1,\sigma}\hat n_I} \big] \nl
&&\quad\quad - \big[ \ave{\hat n_{I-1,I,\sigma}\hat n_J} + \ave{\hat n_{I,I+1,\sigma}\hat n_J} \big] \bigg\}.
\eeqa
Unlike equation \eref{ninjr1cr} for the creation process, this correlator for the annihilation process is not only a function of higher order single-site correlators. Indeed, the correlation of single sites with arcs is evident. The first term on the right indicates correlation if $I$ and $J$ are nearest neighbours. The remaining terms show that site occupancy at site $I$ is correlated with the presence of an arc between $J$ and one of its nearest neighbours, and vice-versa. This makes sense, since the process under consideration here is the annihilation of arcs, which can only occur if an arc is present between nearest-neighbouring sites. It is for this reason that correlation to arc number operators (and not just single-site operators) is observed.

For the annihilation process we obtain the following result for the arc density,
\beqa
\partial_t \ave{\hat n_{I,J,\tilde\sigma}}_{\textnormal{R1,an.}} &=& -h \sum_i \sum_\sigma \;\delta_{I,i}\,\delta_{J,i+1} \,\delta_{\sigma,\tilde\sigma} \;\ave{c_{i,i+1,\sigma}} \nl
&=& -h \;\delta_{<I,J>}\ave{\hat n_{I,J,\tilde\sigma}}. \label{r1bowdens}
\eeqa
Here it is clear that there must be an arc species $\tilde\sigma$ present in order for the \textbf{R1} annihilation process to decrease the arc-density $\hat n_{I,J,\tilde\sigma}$, and this can only happen if $I$ and $J$ are nearest neighbouring sites. 

\begin{figure}[H]
	\centering
		\includegraphics[width=0.4\textwidth]{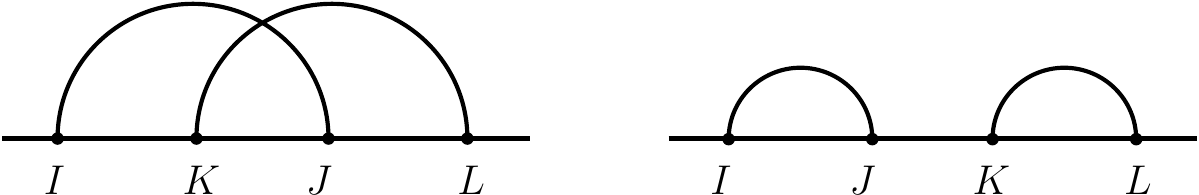}
	\caption{Two different initial conditions. In the first case, it is impossible that dynamics under \textbf{R0} and \textbf{R1} ever result in removal of either arc. In the second case this is not true.}
	\label{fig:r1bow_ch5}
\end{figure}

Lastly we calculate the arc correlator for the annihilation process,
\beqa
\partial_t \ave{\hat n_{I,J,\sigma_1}\,\hat n_{K,L,\sigma_2}}_{\textnormal{R1,an.}} &=& -h\; \Big\{\delta_{<I,J>}\ave{\hat n_{I,J,\sigma_1}\,\hat n_{K,L,\sigma_2}} \nl
&& \;\;\;+ \;\;\delta_{<K,L>}\ave{\hat n_{K,L,\sigma_2}\,\hat n_{I,J,\sigma_1}}\Big\}. \label{r1bowcor}
\eeqa
Since we are only considering the \textbf{R1}-annihilation move, it makes sense that two arcs can only be correlated if one of them could be removed through \textbf{R1}. This is ensured by the Kronecker delta functions and the occupancy number combinations. In equations \eref{r1bowdens} and \eref{r1bowcor} we note an explicit dependence on the initial conditions: if the initial configuration is such that two arc feet at sites $I$ and $J$ can never become nearest neighbours through \textbf{R0} arc diffusion, then this arc cannot be removed. Here it is instructive to consider the two cases in Figure \ref{fig:r1bow_ch5}.

It is thus clear that these dynamics encode the required topology conservation. This makes explicit the dependence on initial conditions (consider, for instance, stochastic evolution of different prime knots such as in Figure \ref{fig:bowdiagexplained}).

To obtain a comprehensive picture of the dynamics it would be instructive to combine the differential equations for the densities or correlators subject to \textit{all} Reidemeister moves considered thus far. This would make explicit the competing effects of annihilation and creation, and may be of particular interest when investigating rate-limiting behaviour of the various dynamical processes.

\subsection{\textbf{R2} and \textbf{R3}: first steps and perspective}
\label{R2doisect}

The Liouvillian for the \textbf{R2}-annihilation process is easy to write down,
\beq
\hat L _{\textnormal{R2,an.}} = h\sum_{i=1}^N \sum_{j=i}^N\sum_{k(i)}\sum_{l(j)}\sum_{\sigma} \bigg\{c_{i,j,\sigma}c_{k,l,\sigma} - \underbrace{c^\dagger_{i,j,\sigma}c_{i,j,\sigma}}_{\hat N_{i,j,\sigma}} \underbrace{c^\dagger_{k,l,\sigma}c_{k,l,\sigma}}_{\hat N_{k,l,\sigma}}\bigg\}.
\label{LR2an}
\eeq
As required, this Liouvillian leaves $\mathcal P$ invariant. Through similar reasoning as in previous sections, it may also be shown to be probability conserving. The first term checks that the precursor state has two neighbouring arcs that that are not present in the ``current'' state. The second term ensures that the current state can only be ``exited'' if the correct configuration is present.

The creation Liouvillian is more difficult to write down, because we need to count the number of arcs completely contained between the sites where we are creating the arc pair. This is required in order to get the orientation of the two strands right --- see Section \ref{R2descript}. This counting procedure is, of course, extremely non-local. In a very rough approximation one could neglect this non-local counting to obtain
\beq
\hat L _{\textnormal{R2,cr.}} \approx g\sum_{i=1}^N \sum_{j=i}^N\sum_{k(i)}\sum_{l(j)}\sum_{\sigma} \bigg\{c^\dagger_{k,l,\sigma}c^\dagger_{i,j,\sigma} - c_{i,j,\sigma}c^\dagger_{i,j,\sigma}\;c_{k,l,\sigma}c^\dagger_{k,l,\sigma} \bigg\},
%\underbrace{c_{i,j,\sigma}c^\dagger_{i,j,\sigma}}_{(1-\hat n_{i,j,\sigma})(1-\hat n_i)(1-\hat n_j)} \underbrace{c_{k,l,\sigma}c^\dagger_{k,l,\sigma}}_{(1-\hat n_{k,l,\sigma})(1-\hat n_k)(1-\hat n_l)}\bigg\}.
\eeq
which is no more difficult to deal with than the \textbf{R2}-annihilation Liouvillian \eref{LR2an}. 

As for \textbf{R0} and \textbf{R1} one can now derive a hierarchy of equations for densities and correlators under \textbf{R2}, although various commutators would be more tricky to calculate. A Liouvillian for the \textbf{R3} move, encoding the rules set out in Section \ref{R3descript}, would be very complex since it requires many different case checks for valid configurations (recall Figure \ref{fig:r3fail}). However, this can be done systematically through our operator formalism.

Extension of the above analyses to the higher order Reidemeister moves would therefore entail the pairwise nearest-neighbouring creation and annihilation processes set out above, and triplet nearest-neighbouring exchange operations subject to configuration checks.

\section{Concluding remarks and outlook}
In this article we have addressed the topological equivalence of knots under the Reidemeister moves. Simple arc diagram representations of knot projections allowed us to recast the Reidemeister moves in terms of dynamical rules on crossings. 

We then established an operator formalism that captures the occupancy restrictions and topological rules on arc diagrams. This necessitated the introduction of composite paulionic operators. Using ideas from reaction-diffusion systems, we showed explicitly how the moves \textbf{R0} and \textbf{R1} may be described in terms of these operators. The Liouvillians for \textbf{R2} were presented with a suggested approximation for handling the non-local aspects of the creation process. In principle the Liouvillian for \textbf{R3} is derivable. In this way, crossings on a single self-entangled loop were described as particles in a quasi-one-dimensional system, subject to motions mapped from the Reidemeister moves. Extension of this description to a complete knot would involve the inclusion of a current term that captures the boundary conditions on arc diagrams.

Using the operator formalism, differential equations for single-site and arc densities and correlators were calculated, subject to stochastic dynamics under \textbf{R0} and \textbf{R1} creation and annihilation. Suggestions were presented for dealing with the higher order Reidemeister moves. We have therefore set out here a systemisation from which differential equations for densities and correlators for crossings of knots are derivable. It is manifest in the evolution of correlation functions for topologically distinct arc diagrams that the dynamics encode rules that leave the knot topology invariant. Some lower order equations are derivable by hand, but this formalism allows in principle for computation of all orders of correlators.

Our approach opens the door to several interesting questions pertaining to the role of topological constraints in rate-limiting behaviour and slowing of dynamics. For instance, one could address time-scales associated with rearrangements, alterations or simplifications of a given knot. Analytical results may be useful for the determination of time-scales in a possible Monte Carlo-type simulation of these dynamics. Through arc correlation functions, for instance, one could investigate (average) time-scales associated with the growth or shrinking of a particular arc. Coupled with appropriate initial conditions, this could be used to study simplification of knots to underlying simpler knots; some concrete suggestions are addressed in Appendix \ref{simchapter}.

Further investigation of our dynamics could allow for the identification of various dynamical regimes of the system and of steady-state solutions. {In particular, it is interesting to ask whether these \textit{purely topological} dynamics may exhibit some significant slowing in certain regimes, as is known for single-file dynamics (see e.g. \cite{schutz2005single}). %Pursuing this question further may provide insights into the topological (rather than interactions-related) origins of glassiness for a single knot (as opposed to distributions of knots). 
Our approach therefore opens new doors to analysing systematically the role of topology conservation in the relaxation or rate-limiting behaviour of crossings subject to Reidemeister moves. It will further be important to study the dependence on initial conditions, since different knot projections may be topologically distinct, and the dynamics conserve topological properties. Therefore various dynamical sectors may arise in dependence on initial conditions. Sensitivity to the rate constants related to the various moves will also need to be considered.}

Our rules are derived from two-dimensional projections of the three-dimensional knot, but we propose that in future this might be an ansatz by which the topology conservation and polymer interactions might be separated. Since we have recast these rules in the setting of particle dynamics, a variety of other techniques is available for further study. {It would be particularly useful to seek a coherent state path integral description as has been done for other reaction-diffusion systems with restricted occupation numbers (e.g. \cite{patzlafftrimper1994,bruneljopa,brunelarxiv}). Such paulionic and fermionic field theories are rather complex. However, with the aid of some clearly-established techniques set out in aforementioned references, we believe that this idea could indeed shed further light on topological dynamics. }

%Due to the paulionic nature of our operators, this would be approached through modified Grassman fields. Although exploratory work in this direction led to some apparent ambiguities for continuum limits, we believe that this idea could shed further light on topological dynamics. 

The descriptions here can be refined in several ways. One could, for instance, make the rate constants for various processes dependent on length-scales to mimic bending energies or to penalise great curvatures. One may also ask whether a particular prime knot underlies some more complex knot. Naturally the discretisation length is relevant in this context, since at most $N$ crossings are possible for $N$ discrete sites on an arc diagram. Drawing on these ideas, we present in Appendix \ref{simchapter} a brief outlook on a possible computational scheme for simulated annealing of knots, based on the rules for dynamics of crossings. As stated, these analytical results could be of use in estimating relevant time-scales for the different dynamical processes in such an algorithm.

{
Our main aim here is to establish a new link between statistical physics tools for studying dynamical regimes of reaction-diffusion systems, and the question of the role of topology conservation in dynamics. As mentioned, our formalism may have limitations regarding tractability, and the connection to physical degrees of freedom (for instance those of a three-dimensional polymer knot) will certainly require careful consideration. However, we hope that this link, which is amenable to a wealth of calculational techniques, provides some fresh perspectives.
}

\appendix
\section{Thoughts towards simulated annealing of knots}
\label{simchapter}

%Topics to cover
%\bi
%\item Measures of knot complexity
%\bi
%\item Include that various quantities are often studied as a function of knot complexity.
%\item Reference to Nechaev:  knot complexity i.t.o. algebraic invariants; limit distributions for this in terms of braid and locally free groups.
%\item Crossing number
%\item Energetic considerations: motivate that global energy minimum is an indicator of the simplest knot form.
%\bi
%\item Moffat's fluid mechanical things
%\item Electrostatic repulsion
%\item Bending energy
%\item Thickest rope / inflation
%\ei
%\ei
%\item Untangling and annealing knots
%\bi
%\item Algorithms:
%\bi
%\item Generating random knots
%\item Sampling RWs
%\item Simulated annealing: the various algorithms. Avoiding local minima.
%\item BCASDKF algorithm
%\item Lattice knots, Van Rensburg etc
%\ei
%\ei
%\ei

It is interesting to ask how one may reduce or simplify a given knot to its ``simplest form''. For instance, some complex knotted structure could perhaps be untangled and simplified to yield the underlying topologically equivalent prime knot. {Indeed, analytical approaches to knot and braid reduction, based on algebraic techniques for reduction of words, have been considered (e.g. \cite{nechaev1996,dynnikov}).} How then does one quantify complexity of a knot? The crossing number has featured extensively here, for instance in the study of topological effects in polymer dynamics \cite{quake1994}, the estimation of the number of knots with $n$ crossings through path integral techniques \cite{kholodenko1996} and for analysing complexity of polygonal lattice knots \cite{diao1998}. The crossing number is also used in computer simulations addressing the relation of knot complexity and knotting probability \cite{shimamura2002}. Other measures of knot complexity include powers of some algebraic invariants, applied in studying knot entropies in the setting of braids and locally free groups \cite{nechaev1996}. 

A very physical approach to finding a ``simplest configuration'' of a knot is through minimisation of suitably defined knot energies. Indeed, energy spectra for knots have been defined through fluid mechanical techniques \cite{moffatt1990}, and knot simplification has been studied in terms of electrostatic repulsion of equidistantly spaced charges along knots \cite{fukuhara1988}. Extensions to this include bending energies \cite{ohara1991} and M\"obius knot energies defined in terms of integrals of curves \cite{freedman1994}. It is even possible to obtain a minimal configuration of knots that is an invariant of knot type through global minimisation of knot energies based on the total curvature and electrostatic-type interactions \cite{buck1993}.

Another physical idea is to ``inflate'' a knot to drive its geometric simplification \cite{grosberg1996inflate}. Various definitions of knot thickness \cite{diao1999} and rope length of knots \cite{cantarella2002} have been applied to knot classification and simplification, and the relation of knot thickness to crossing number has been studied\cite{buck1999}.

Since many of the above-mentioned ideas are very difficult to deal with analytically, the simplification of knots has also been studied extensively through computational / algorithmic techniques. Indeed, it is this question which we address below. Such procedures require two ingredients, namely a measure of knot complexity (e.g., the crossing number), and a driving process for the simplification / reduction (e.g., minimisation of knot energy). We briefly outline here how these concepts have been employed in simulations. Thereafter we suggest how our labelling scheme and dynamical rules could be employed to yield a simple algorithm for knot simplifications.

\subsection{Algorithmic untangling knots: a brief overview}

One key motivation for finding algorithmic prescriptions to simplify knots is the study of probability distributions of randomly generated knots \cite{shimamura2002} and to test for the equivalence of given knots. The latter question is particularly challenging, since it would otherwise involve testing for the existence of some sequence of Reidemeister moves that relates two knots (computationally an open problem), or calculating some knot invariant (an incomplete test for knot equivalence) \cite{huang1996}. 

Knot simplification simulations typically involve the improvement of some cost function related to the knot complexity. It has, for instance, been suggested to evolve knots iteratively along the gradient of some chosen energy function deterministically (e.g. \cite{fukuhara1988}) or stochastically (e.g. \cite{simon1994}) through the introduction of small perturbations that are accepted if they affect the cost function favourably. Such techniques, however, are sensitive to getting stuck in local stationary states \cite{huang1996}. To avoid this issue, the problem has been tackled through simulated annealing \cite{huang1996,ligocki1994}. This approach involves the occasional acceptance of configurational changes that evolve the system \textit{against} the gradient of the cost function. The acceptance of such ``uphill'' perturbations is often related to a temperature parameter, so that initially they are more likely to occur. The system is then gradually ``cooled off'' with time, and uphill perturbations become less likely. (The applicability of such Metropolis-type algorithms to annealing and optimisation problems in statistical physics has been demonstrated extensively; see, for instance, the highly-cited article of Kirkpatrick \etal \cite{kirkpatrick1983}.) Various uphill perturbation methods have been used, including placing point charges near the (charged) knot and allowing it to evolve under electrostatic forces, tightening or loosening various parts of the knot \cite{huang1996}, and perturbing the vertices of a piecewise linear curve representation of the knot \cite{ligocki1994}. More recently, a considerable improvement in computation times has been achieved through algorithms that combine energy minimisation and tree-based probabilistic planning \cite{ladd2004}.

%Local deformations and perturbations have also been used to investigate the equivalence of lattice polygons \cite{vanrensburg1991}. \textit{Tie this in better.}

Several of the techniques addressed above are computationally intensive. We shall therefore outline here some ideas for a simple annealing algorithm contingent on our dynamical rules.

\subsection{Simulated annealing: suggestions based on arc diagrams or the Gauss code}

The steps of a suggested algorithm for simulated annealing of knots are discussed below.

\textbf{A. Generation of a random knot}

If this step is required, a random knot can be generated using pivot-type algorithms \cite{madras1988} or closures of random walks (see \cite{virnau2006} and references therein).

\textbf{B. Projection of the knot}

The random knot can now be projected to yield an arc diagram according to Section\ref{knotchapt}. Alternatively, existing tools for finding the Gauss code of the random knot could be augmented to further record distances between consecutive crossings in the projection. (A possible point of departure is the \textit{Mathematica} package \verb+KnotTheory+ \cite{imafuji2002,barmathematica}.) 

\textbf{C. Cost function}

For the purpose of driving the evolution of the annealing algorithm, we suggested a cost function with two core attributes. Firstly, a large crossing number should be penalised so that simplification (i.e., reduction of the crossing number) is favoured. Secondly, free loops should be penalised. Free loops on the random knot are any simple loops that could be removed through \textbf{R1} annihilation; see Figure \ref{fig:r1bow_ch5}. This aspect of the cost function would depend on the arc-length of such a loop (obtainable from the arc diagram), so that small loops are heavily penalised. 

The total cost function for the complete knot is then the sum of contributions of all trivial loops together with the part that depends on the  crossing number.

\textbf{D. Simulated annealing}

We split the simulation processes into several classes.

\textbf{D1. \textit{Random stochastic evolution of the knot}}

This is done according to the Reidemeister moves \textbf{R0} and \textbf{R3} (these govern ``diffusion-type'' processes), and annihilation moves \textbf{R1} and \textbf{R2} (these govern ``simplification-type'' processes that reduce the crossing number). Begin by selecting an occupied site on the arc diagram and allowing that arc-foot to diffuse under the following provisos.
\bi
\item \textbf{R0} allows the segments of the knot to ``diffuse'' relative to each other. Diffusion may only occur to \textit{empty} nearest-neighbouring sites.
\item If the selected arc foot is part of a primitive loop, then there is a probability of removing this loop through an \textbf{R1} annihilation step, related to the net reduction of the cost function.
\item Should the selected arc foot be adjacent to an occupied site, test for the following:
\bi
\item If a valid \textbf{R3} configuration is present, allow execution of \textbf{R3}. If not, the particle may not diffuse.
\item If the configuration allows for an annihilation move of the \textbf{R2}-type, allow this move with a certain probability (depending on the improvement of the cost function through reduction of the crossing number).
\ei
\ei
Every diffusion step should be such that it either reduces the cost function or leaves it unaltered.

\textbf{D2. \textit{Perturbations}}

The second class of simulation processes involves the introduction of new crossings at empty sites on the arc diagram. This can be done in two ways.
\bi
\item Introduce new primitive loops into the arc diagram through an \textbf{R1} creation step. This is an ``uphill'' perturbation: both the crossing number aspect \textit{and} the primitive loop aspect of the cost function are affected against the simplification gradient. Likelihood of such an uphill perturbation is then related to a temperature parameter in the system (see \textbf{D3}). If a new primitive loop has been introduced into the arc diagram, there is a chance of it ``diffusing'' into the rest of the knot through the stochastic evolution; see \textbf{D1}.
\item Introduce crossing pairs through an through an \textbf{R2} creation step. The cost function is affected through the resulting increase in crossing number, again against the simplification gradient. These crossings, too, may diffuse into the rest of the knot through the stochastic evolution; see \textbf{D1}.
\ei

\textbf{D3. Cooling}

The temperature parameter (i.e., the likelihood of uphill perturbations) is then reduced as simulation time progresses, inducing the system to settle to a minimal configuration.

\vspace{0.5cm} \noindent
The core ingredient to our algorithm is therefore a cost function that penalises trivial loops of the knot and favours smaller crossing numbers. Stochastic evolution and perturbations of the knot may were suggested, based on the Reidemeister moves. The introduction of uphill perturbations implies that our algorithm should be less sensitive to local minima of the cost function. The manipulations are on a quasi-one-dimensional representation of the knot, and therefore the computational cost of the algorithm would be minimal. Such an algorithm could be particularly useful in investigating the importance of the individual Reidemeister moves in a purely topological dynamical setting, for instance through relative weighting of rates that govern the different processes. In this way one could study rate-limiting effects in our topological dynamics, possibly as an indicator of topologically-induced glassy behaviour. The relevance and usefulness of our algorithm remains to be tested. It is not clear whether the projection procedure is necessarily easy, or whether sufficiently much information about the real-space configuration of the knot is maintained after the projection. Consequently other types of cost functions could be considered.

%%%%%%%%%%%%%%%%%%%%%%%%%%%%%%%%%%%%%%%%%%%%%%%%%%%%%%%%%%%%%%%%%%%%%%%%%%%%%%%%%%%%%%%%%%%%%%%%%%%%%%%%%%%%
%%%%%%%%%%%%%%%%%%%%%%%%%%%%%%%%%%%%%%%%%%%%%%%%%%%%%%%%%%%%%%%%%%%%%%%%%%%%%%%%%%%%%%%%%%%%%%%%%%%%%%%%%%%%
%%%%%%%%%%%%%%%%%%%%%%%%%%%%%%%%%%%%%%%%%%%%%%%%%%%%%%%%%%%%%%%%%%%%%%%%%%%%%%%%%%%%%%%%%%%%%%%%%%%%%%%%%%%%
%%%%%%%%%%%%%%%%%%%%%%%%%%%%%%%%%%%%%%%%%%%%%%%%%%%%%%%%%%%%%%%%%%%%%%%%%%%%%%%%%%%%%%%%%%%%%%%%%%%%%%%%%%%%

\begin{acknowledgements}
C.M.R. would like to thank Dr. J.N. Kriel for several useful discussions, and acknowledges financial support from the Wilhelm Frank Bursary Trust and Stellenbosch University.
\end{acknowledgements}

\bibliographystyle{unsrt}       % APS-like style for physics
\bibliography{references}   % name your BibTeX data base

\end{document}